\documentclass[journal=ancac3,manuscript=article]{achemso}

\usepackage{adjustbox}

\usepackage{verbatim}
\usepackage[colorlinks=false,bookmarks,pdftitle={},pdfauthor={},pdfsubject={},pdfkeywords={},breaklinks]{hyperref}

\usepackage[usenames]{color}
\usepackage{amssymb,mathbbol,amsmath,amsbsy}
\usepackage{amsfonts}
\usepackage{dsfont}
\usepackage{amsmath}

\usepackage[utf8]{inputenc}
\usepackage{graphicx}

\usepackage{multirow}

\usepackage{booktabs}
\usepackage[dvipsnames]{xcolor}

\usepackage{booktabs}
\usepackage{multirow}

\usepackage{pifont}
\usepackage{bm}
\usepackage[mathlines]{lineno}

\graphicspath{{figures/}}

\newcommand{\cmark}{\ding{51}}
\newcommand{\xmark}{\ding{55}}

\providecommand{\onecolumngrid}{}

\newenvironment{ruledtabular}{}{}

\newcommand{\trans}{\mathcal{T}}
\newcommand{\kB}{k_\mathrm{B}}

\newcommand{\kappaC}{\kappa_\mathrm{C}}
\newcommand{\kappaQ}{\kappa_\mathrm{Q}}
\newcommand{\Ang}{\mathrm{\AA}}
\newcommand{\WmK}{\mathrm{W}\,\mathrm{m}^{-1}\mathrm{K}^{-1}}
\newcommand{\cm}{\mathrm{cm}^{-1}}

\newcommand{\PR}{\mathrm{PR}}

\newcommand{\supplement}{Supporting Information}

\newcommand{\baslik}{Quantum Statistics and Structural Topology Govern Thermal Transport in Two-Dimensional Monolayer Amorphous Carbon}

\title{\baslik}

\author{G. Kurt}
\affiliation{
Department of Materials Science and Engineering, \.Izmir Institute of Technology, G\"ulbah\c{c}e Kamp\"us\"u, 35430 Urla, \.Izmir, T\"urkiye}

\author{H. Sevin\c{c}li}
\email{sevincli@fen.bilkent.edu.tr}
\affiliation{Department of Physics, Bilkent University, 06800 Ankara, T\"urkiye}

\begin{document}

\maketitle

\begin{abstract}

    We investigate the quantum thermal conductivity (TC) of two-dimensional monolayer amorphous carbon (MAC).
    We employ three distinct amorphization algorithms to generate various possible MAC configurations, ranging from Zachariasen-type continuous random networks to nanocrystallites embedded in random networks.
    The local bond order parameter, $q_3$, is used to quantify the amorphousness of the structures, and TC is computed as functions of $q_3$ and temperature.
    This framework enables us to assess how structural topology, degree of amorphization, and quantum statistics contribute to heat conduction in a two-dimensional amorphous solid.
    At room temperature, TC values are predicted to range from 3.5 to 10~$\WmK$, within the range of recent experimental measurements.
    Analysis of vibrational modes reveals that, while the modes of these 2D amorphous structures fall into the usual categories, namely, propagons, {diffusons, and locons}, their polarization characteristics display distinct behavior.
	Owing to the fully quantum mechanical framework, we examine both low- and high-temperature characteristics of this 2D amorphous system.
 	By examining the classical limit, we show that classical treatments substantially overestimate the TC of MAC; namely, the quantum TC is less than half of the classical value at room temperature and up to nearly an order of magnitude lower at low temperatures.

\end{abstract}

\section{Introduction}

Two-dimensional amorphous materials have recently emerged as an important class of systems with promising applications across diverse fields~\cite{toh_nature_2020}. In contrast to their crystalline counterparts, which have been studied far more extensively, amorphous 2D systems remain comparatively scarce and much less understood. Against this background, two-dimensional amorphous systems, and in particular MAC, offer a uniquely accessible platform to probe the physics of strongly disordered materials and to explore fundamental structure-property relationships~\cite{kotakoski_prl_2011,tian_acsnano_2023,wang_adfm_2025,tian_nature_2023,wang_prx_2025}.

Thermal conductivity (TC) is a particularly striking example of this knowledge gap. The TC of crystalline materials is well established, especially in three-dimensional bulk systems at room temperature and above, where semiclassical approaches such as the Boltzmann transport equation combined with first-principles force constants yield highly accurate results. By contrast, the TC of amorphous materials remains under debate, particularly at low temperatures where their behavior cannot be captured by semiclassical theories~\cite{zhu-2016-nanoletters,antidormi-2020-2dmat,wang_prx_2025}.
Lacking long-range order, amorphous solids display substantially lower TC than their crystalline counterparts and exhibit fundamentally distinct temperature dependence, governed primarily by elastic rather than anharmonic processes. Yet, despite their significance, little is known about the thermal transport of two-dimensional amorphous systems, leaving central questions in both fundamental science and potential applications unanswered.

In this work, we systematically investigate different network topologies and structural motifs with various degrees of amorphization in MAC to uncover their impact on thermal transport, by including structures generated using distinct amorphization protocols and spanning a wide temperature range. We employ a fully quantum mechanical framework of Landauer formalism based on Green’s function techniques, which we further extend to consistently recover the classical (Boltzmann) limit. This approach provides a consistent description of amorphous heat transport by computing transmission amplitudes directly, requires no a posteriori quantum corrections, and avoids invoking phonon gas models.

Unlike the usual case, where quantum effects are important only at very low temperatures, in MAC they remain significant across a much broader temperature range. We show that quantum TC is less than half the classical value at room temperature and by nearly an order of magnitude at low temperatures. At the same time, comparisons among 3C-GM, 3C, and NC@RN structures show that structural topology and the degree of amorphization determine the underlying transmission spectra. This establishes MAC as a model platform for assessing the roles of topology-driven vibrational scattering and quantum statistics in heat conduction in two-dimensional amorphous materials.

Details on physical effects such as buckling, further analysis of the vibrational spectrum, inverse participation ratios, dynamical structure factors, and computational methods such as potential parametrization, amorphization schemes, the order parameter, partitioning, and Green function techniques are given in the \supplement.

\section{Results and Discussion}
\label{sec:results}

We first establish how the different amorphization protocols modify the structural topology of MAC, then connect these structural features to vibrational-mode character and finally to thermal conductivity.

\subsection{MAC Structures}

We first investigate the structural aspects.
To model MAC, we combine empirical potentials with Monte Carlo–based structure generation schemes. The Adaptive Intermolecular Reactive Empirical Bond Order (AIREBO) potential is used to generate initial configurations because of its proven reliability in reproducing structural features of amorphous carbons, while the re-optimized Tersoff potential is adopted for subsequent relaxation and force constant extraction, owing to its success in describing vibrational spectra and thermal transport properties.\cite{airebopot,tersoff_prb_1988,tersoff_prl_1988,lindsay-2010-prb} All structural relaxations and total energy evaluations were performed using the LAMMPS package,\cite{LAMMPS} with all degrees of freedom fully relaxed until atomic forces fell below a $10^{-3}$~eV/$\Ang$ threshold. Further details of the relaxation protocols and potential parametrization are provided in the \supplement.

Amorphous configurations are constructed through an energy-driven kinetic Monte Carlo  approach, which generates bond rotations followed by local relaxation. This method allows us to generate three representative structural motifs, two of which are Zachariasen-type continuous random networks (CRNs), and one hosting nano-crystallite islands in otherwise  random networks, to be referred to as NC@RN.
{In CRN structures, all carbons are three-coordinated, and they are generated using the 3C (three-coordinated) or the 3C-geometric-modeling (3C-GM) algorithms, which follow a crystalline to amorphous approach~\cite{kumar-2012-jopcm}}.
In 3C-GM structures, rings with less than five members are not allowed, whereas in 3C there is no restriction on the ring size.
The NC@RN structures, on the other hand, are generated by following a random-to-ordered approach~\cite{zhuang-2016-pccp, toh_nature_2020}, and there are no constraints about the coordination number or the ring sizes.
Hence we refer to the algorithm as the non-constrained (nC) scheme.

Representative continuous random network structures generated using the 3C-GM and 3C algorithms, together with an NC@RN structure generated using the nC algorithm, are shown in Fig.~\ref{fig:structural}a from left to right: 3C-GM, 3C, and nC.
A table summarizing the structural types and their fundamental characteristics can be found in Table~\ref{table:structuraltypes} of \supplement.

The amorphousness of structures is quantified with the local bond order parameter, which yields a measure of the structural symmetry around an atom.~\cite{steinhardt-1983-prb,antidormi-2020-2dmat}
For the $i^{th}$ atom, it is defined as
$q_{lm}{=}N_i^{-1} \sum_{j\epsilon\mathcal{S}_i} Y_l^m(\boldsymbol{r}_{ij})$,
where $\mathcal{S}_i$ is the set consisting of nearest-neighbors to the $i^{th}$ atom, and $N_i$ is the number of elements of $\mathcal{S}_i$.
A neighboring atom is identified as a first-nearest-neighbor if it is closer than 2.3~\AA.

$ Y_l^m(\mathbf{r}_{ij})$ is the spherical harmonic and $\mathbf{r}_{ij}$ is the vector from the $i^{th}$ atom to the $j^{th}$ atom.
The local bond order parameter for each atom is calculated as
\begin{equation}
    \Tilde{q}_l(i) = \frac{1}{N_i}  \sum\limits_{j\epsilon\mathcal{S}_i} \mathbf{q}_l(i) \cdot \mathbf{q}_l(j).
\end{equation}
The system's order parameter is determined by averaging $\Tilde{q}_l(i)$ over all atoms. Here, because of the hexagonal structure of the crystalline counterpart, $q_3$ is the relevant order parameter. This parameter takes values between 0 and 1 and equals 1 for a perfect crystal~\cite{eslami-201-joctac}.

The colors of the atoms in Fig.~\ref{fig:structural}(a) represent their $q_3$ values, and the average $q_3$ values of the structures are almost the same, $q_3 = 0.60$.

The radial distribution function (RDF) provides a clear fingerprint of amorphization (Fig.~\ref{fig:structural}b). At short distances, the first-nearest-neighbor peaks broaden as $q_3$ decreases, while beyond 5~$\Ang$ the peaks lose coherence, signaling the absence of long-range order. Bond length and angle distributions follow the same trend, spreading around the crystalline graphene values of 1.43~$\Ang$ and 120$^\mathrm{o}$. In NC@RN structures with low $q_3$, a secondary short-bond peak near 1.33~$\Ang$ arises from $sp$-hybridized atoms and three-membered rings, whereas NC@RN structures with higher $q_3$ exhibit sharper features due to crystalline inclusions. These signatures align with experimental observations of MAC~\cite{toh_nature_2020}.

Ring statistics further distinguish the structural motifs (Fig.~\ref{fig:structural}e). In CRNs, reduced hexagon density with decreasing $q_3$ is offset mainly by pentagons and heptagons. NC@RN configurations, however, also produce large rings with nine or more atoms, consistent with the coexistence of crystalline clusters and void-like regions. Between the two CRN schemes, 3C retains more hexagons, while 3C-GM favors heptagons under comparable $q_3$ values due to its stricter geometric constraints. In both cases, hexagons are typically isolated at defect boundaries, reflecting the nature of continuous random networks.

As expected, MAC structures exhibit significant buckling, and it becomes more pronounced with increasing disorder. For example, in highly disordered 3C-GM structures characterized by $q_{3} = 0.55$, the average out-of-plane displacement of the atoms reaches values as high as 4~\AA{} (Fig.~\ref{sfig:buckling}). Such large vertical deviations from the basal plane clearly indicate substantial structural corrugation in these systems, and are known to affect thermal transport substantially~\cite{sevincli_apl_2014}.
CRN and NC@RN structures exhibit distinct buckling behaviors, reflecting differences in their local bonding environments and structural organization. To quantify these differences, we analyze the orientation of bonds with respect to the basal plane. In NC@RN structures, the average angle between the bonds and the basal plane increases from $9.2^\circ$ to $12.3^\circ$ as $q_{3}$ decreases from 0.70 to 0.55, indicating a moderate enhancement of out-of-plane distortion with increasing disorder.
In contrast, the corresponding bond orientation angles in 3C structures are considerably larger. Specifically, these angles are $14.5^\circ$ at $q_{3} = 0.70$ and increase to $19^\circ$ at $q_{3} = 0.55$, demonstrating a substantially stronger deviation from planarity compared to NC@RN systems at similar levels of disorder (see Fig.~\ref{sfig:bucling_bondorientation} for details).

\subsection{Vibrational Mode Analysis}

\begin{figure}[t]
	\centering
	\includegraphics[width=.75\textwidth]{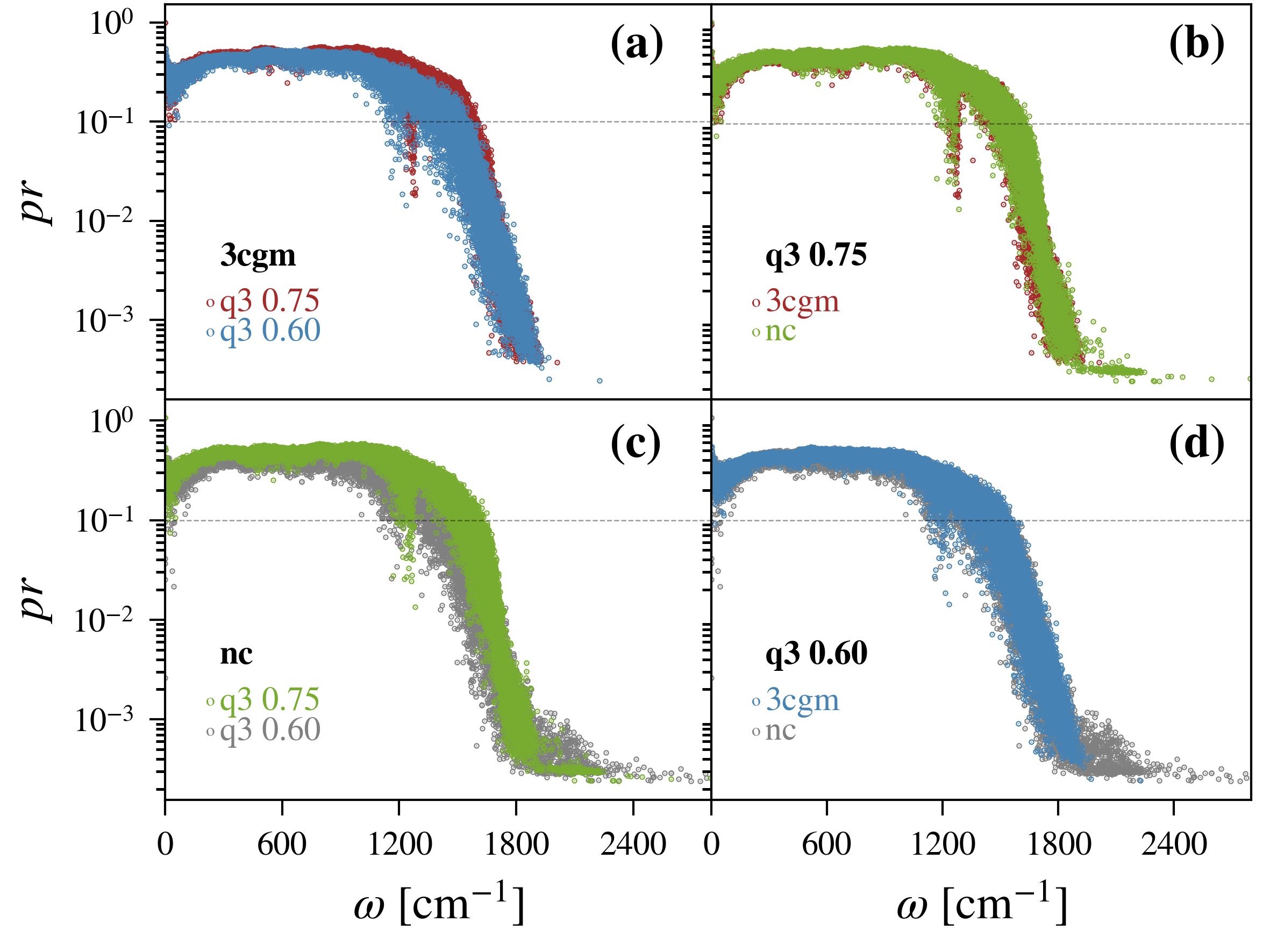}
	\caption{
    Mode participation ratios (PRs) for 3C-GM and NC@RN structures with order parameter values of $q_3=0.60$ and $q_3=0.75$.
	}
	\label{fig:pr}
\end{figure}

The nature of vibrational modes is important in determining thermal conductivity.
We investigate them by analyzing their participation ratios, polarization spheres, Fourier transforms, and phase quotients.
The mode participation ratio (PR) quantifies the number of atoms that significantly participate in a given normal mode, and it is defined as
\begin{equation}
    \label{eqn:PR}
	\mathrm{PR}_m=
    \frac{1}{N}
    \frac{
	\big(
    \sum_{i\alpha}
	  \epsilon_{m,i\alpha}^2
	\big)^2
    }{
	\sum_{i}
    \left(
    \sum_{\alpha} \epsilon_{m,i\alpha}^2
    \right)^2
    }
    ,
\end{equation}

where $\epsilon_m$ is the polarization vector of mode $m$,
$i$ and $\alpha$ denote the atom and the Cartesian indices, and $N$ is the number of atoms.
In crystalline materials, vibrational modes are delocalized and their $\PR$ values are of order unity. In contrast, for localized modes in disordered systems,
$\PR$ scales as $1/N$.
PRs of MAC structures are shown in Fig.~\ref{fig:pr}. In the left panels, we compare MACs with different $q_3$ values but the same structural type, 3C-GM or NC@RN.
Higher $q_3$ results in higher $\PR$, as expected.
Adopting a threshold value of $\PR=0.1$ for distinguishing locons from extendons, we identify the mobility edges to be realized at similar frequencies, independent of the structures being CRN or NC@RN. Namely, the minimum and maximum frequencies
that cross the $\PR=0.1$ limit are the same for $q_3=0.75$, whether the structure is CRN or NC@RN (Fig.~\ref{fig:pr}(b)). The same holds for $q_3=0.6$, as seen in Fig.~\ref{fig:pr}(d).
A comparison of structures with different $q_3$ values reveals a clear red-shift of the mobility edge with increasing $q_3$, for both CRNs and NC@RNs,

Fig.~\ref{fig:pr}(a,~c).

NC@RNs also exhibit extremely localized vibrational modes that lie beyond the maximum vibrational frequency supported by graphene, as seen in Fig.~\ref{fig:pr}(c).
These modes are due to low-coordinated carbons in these structures.

\begin{figure}[t]
	\flushleft (a)\\
	\centering
    \includegraphics[width=.7\textwidth]{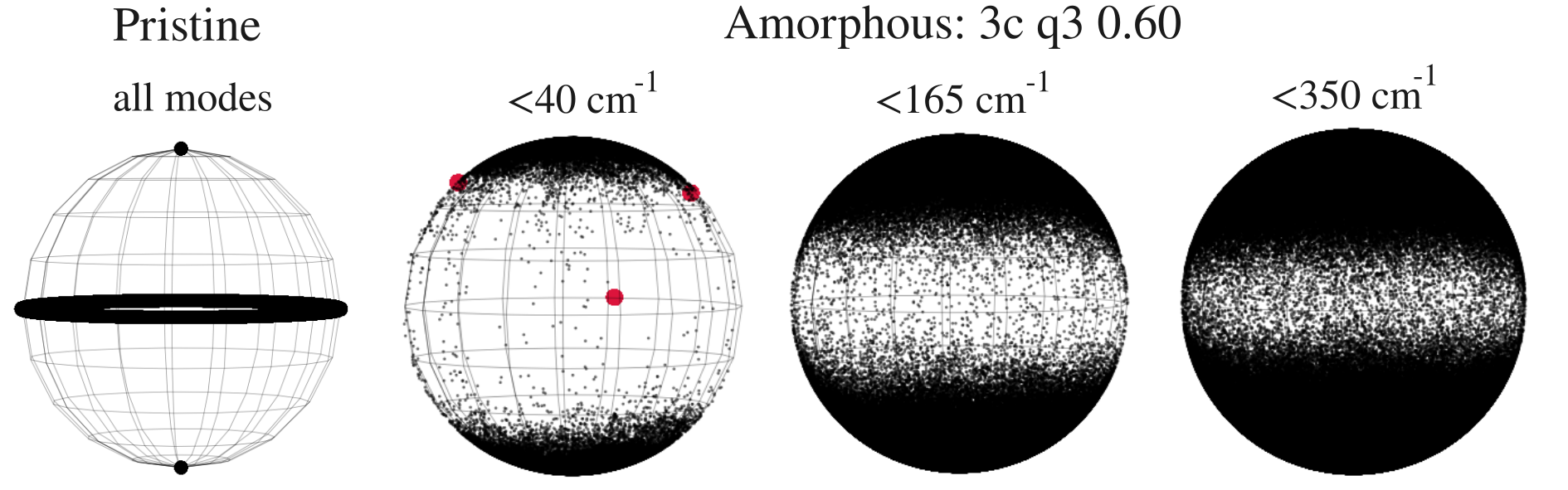}

    \flushleft (b)\\
    \centering
	\includegraphics[width=.7\textwidth]{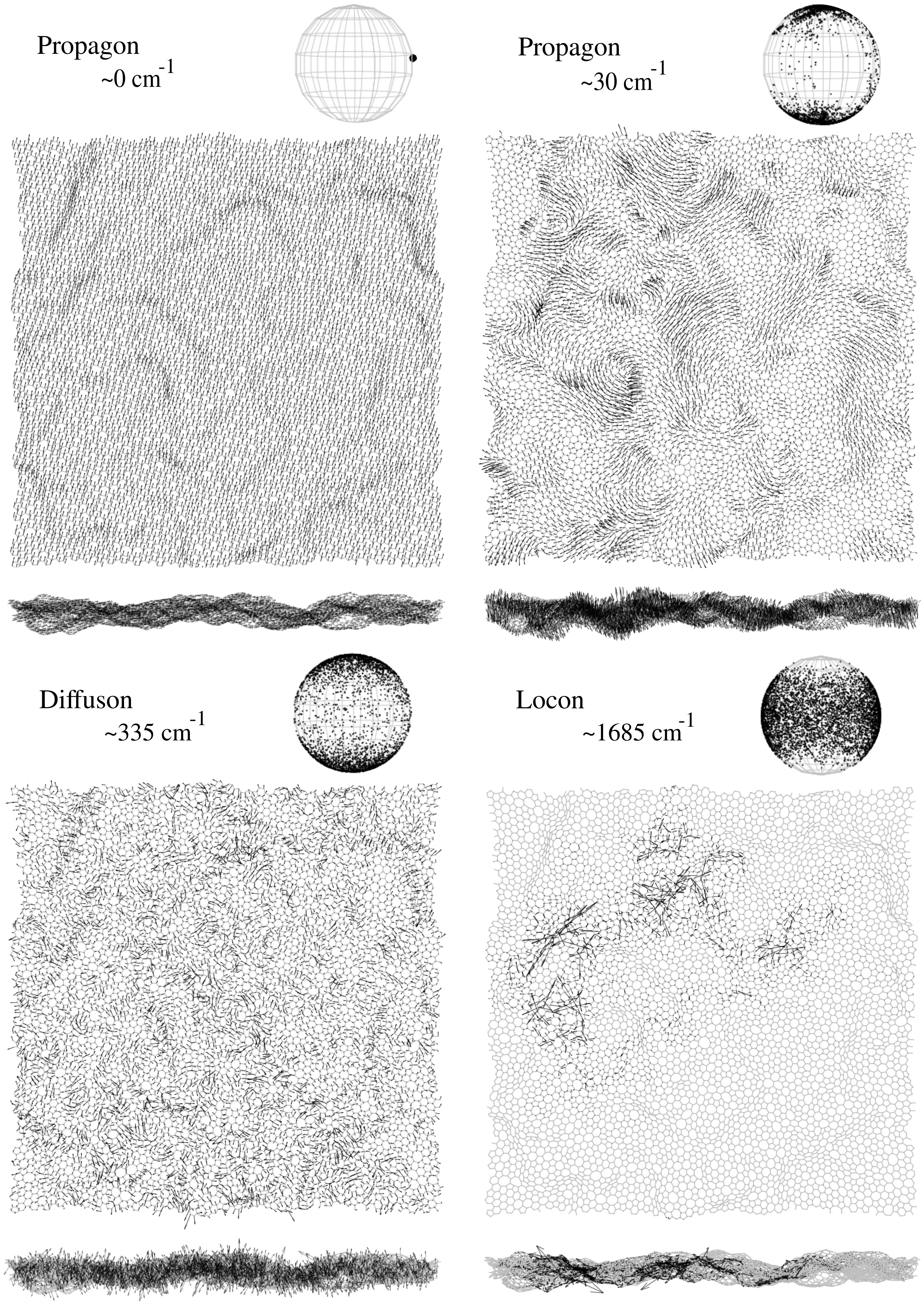}
	\caption{(a) Polarization spheres of 2D crystalline graphene (all modes) and amorphous carbon (certain frequency ranges) contrast the directionality of crystalline and amorphous structures in 2D.
    (b) Atomic displacements and corresponding polarizations of four selected modes illustrate the distinctive features of propagons, diffusons and locons.
	}\label{fig:polarization}
\end{figure}

In addition to $\PR$, the modal polarization sphere provides useful information on the nature of modes in amorphous materials (Fig.~\ref{fig:polarization}).
In pristine graphene, the vibrational modes are either in-plane or out-of-plane, so their polarization is either on the equator or the poles.
With increased disorder, atomic displacements associated with a given mode $m$ become increasingly randomized, losing the directional coherence.
The polarization sphere representation is constructed from the normalized eigenvector components $\epsilon_{m,i\alpha}$.
Atomic polarizations manifest a transition from clustered, directional vectors in low-frequency modes to an isotropic
distribution at higher frequencies. Propagons possess a well-defined polarization and often display an acoustic character with dominant out-of-plane contributions, signifying propagating waves, and are typically observed in the lower frequency range.
In Fig.~\ref{fig:polarization}(a), polarization sphere of pristine graphene, as well as propagons, diffusons and locons in MAC can be distinguished from their polarizations.
Diffusons lack a meaningful wave vector and exhibit spatially extended but randomized vibrational fields, whereas locons are spatially localized.

At low frequencies ($\omega<40\;\cm$), the polarization remains relatively well-defined, with atomic motions predominantly aligned along the $z$-direction, reminiscent of the out-of-plane modes in graphene. However, above this frequency, atomic motions start to deviate from specific orientations, showing diffusional character and the absence of a wave-vector. At higher frequencies, the polarization directions become uniformly distributed, implying a complete loss of wave-like behavior.
Fig.~\ref{fig:polarization}(b) illustrates the top and side views of MAC structures with atomic displacements and corresponding polarization spheres of four selected modes, two of which are propagons (with frequencies of ${\approx}0$~$\cm$ and $30$~$\cm$), a diffuson ($335$~$\cm$) and a locon ($1685$~$\cm$).
The atomic structure is plotted in gray, and the black arrows show the direction and magnitude of each atom's motion associated with the selected mode.
For the lowest frequency propagon, all atoms move  in the same direction and amplitude, and this corresponds to an isolated dot on the polarization sphere.
Such a single direction is absent for the 30~$\cm$ propagon, but a correlated motion of nearby atoms still exists and the polarization sphere is populated close to the poles.
Vortex-like displacement patterns
and unequal magnitudes in atomic displacements are among the distinctive features of this propagon.
At 335~$\cm$, the atomic motions are quite random in both magnitude and direction, which is also evident in the polarization sphere. The local correlations as observed for the 30~$\cm$ propagon are diminished at 335~$\cm$.
The locon (1685~$\cm$), on the other hand, consists of atomic displacements confined to small regions with most of the atoms remaining nearly immobile while others vibrate in random directions with very high amplitudes.

\subsection{Thermal Conductivity}

In contrast to charge transport, which is restricted to electronic
states near the Fermi energy, vibrational heat transport involves
the entire frequency spectrum and different spectral regions
exhibit distinct transport regimes. In crystalline materials, heat
is carried by phonons: low-frequency phonons propagate
quasi-ballistically, mid-frequency phonons diffuse,
high-frequency modes are often localized, and hence heat is carried mostly by acoustic phonons.
The same applies to  propagons, diffusons, and locons in amorphous solids~\cite{allen_prb_1993,allen_pmb_1999}.
The predominance of low-frequency modes arises not only from the short mean free paths of the high-frequency modes, but also because of quantum statistical effects.
Quantum statistics becomes significant at temperatures
below \(\hbar\omega_{\mathrm{max}}/\kB\), which corresponds to
approximately 2300~K for 2D carbon. A fully quantum-mechanical
treatment of vibrational heat transport is therefore essential
for MAC.

We employ the Landauer formalism, which treats ballistic, diffusive, and localized transport regimes on equal footing without \textit{a priori} assumptions about the transport regimes.
The method combines transmission probabilities with Bose–Einstein statistics and provides a consistent description of vibrational heat transport across the entire frequency spectrum.
The vibrational thermal conductivity is expressed as
\begin{equation}
	\kappa(T)=\frac{\kB L}{2\pi A}
	\int d\omega\,
	p(\omega,T)\,
	\trans(\omega),
\end{equation}
where \(\trans(\omega)\) denotes the transmission function, obtained
from atomistic Green's function calculations~\cite{rego_prl_1998,sevincli-jopcm-2019},
and \(p(\omega,T)\) is the quantum-statistical weight function. Further details are included in the \supplement.
For amorphous materials, anharmonic effects can be neglected to a
good approximation~\cite{lv-2016-apl,allen_pmb_1999}, which is also supported for MAC experimentally up to 450~K~\cite{wang_prx_2025}.
Within the Landauer formalism combined with atomistic Green's functions and Bose–Einstein statistics, elastic scattering induced by structural disorder is treated explicitly and without approximation within the harmonic framework.

\begin{table*}
        \centering
        \begin{ruledtabular}
        \begin{tabular}{c ccc ccc ccc}
            \multirow{2}{*}{Local bond order}        &
                \multicolumn{3}{c}{$\kappaC$ (W m$^{-1}$K$^{-1}$)} &
                \multicolumn{3}{c}{$\kappaQ$ (W m$^{-1}$K$^{-1}$)} &
                \multicolumn{3}{c}{$\kappaC/\kappaQ$} \\
            \cmidrule(lr){2-4} \cmidrule(lr){5-7} \cmidrule(lr){8-10}
            parameter ($q_3$) &
                3C-GM   & 3C    & nC &
                3C-GM   & 3C    & nC &
                3C-GM   & 3C    & nC \\
            \midrule
            0.70 &
                18.48   & 18.93 & 14.08 &
                9.48    & 9.46  & 6.98  &
                1.95    & 1.96  & 2.02\\
            0.55 &
                10.77   & 10.88 & 9.58 &
                6.04    & 5.92  & 4.97 &
                1.79    & 1.78  & 1.93 \\
            0.40 &
                6.86    & 6.74  & 6.74  &
                4.02    & 3.79  & 3.55  &
                1.74    & 1.74  & 1.89\\
        \end{tabular}
        \end{ruledtabular}
        \caption{Room temperature thermal conductivities of monolayer amorphous carbon for various $q_3$ values. The classical ($\kappaC$) and quantum mechanical ($\kappaQ$) results, and their ratio ($\kappaC/\kappaQ$) are tabulated.}
        \label{table:conductivity}
\end{table*}

\begin{figure*}
	\includegraphics[width=\textwidth]{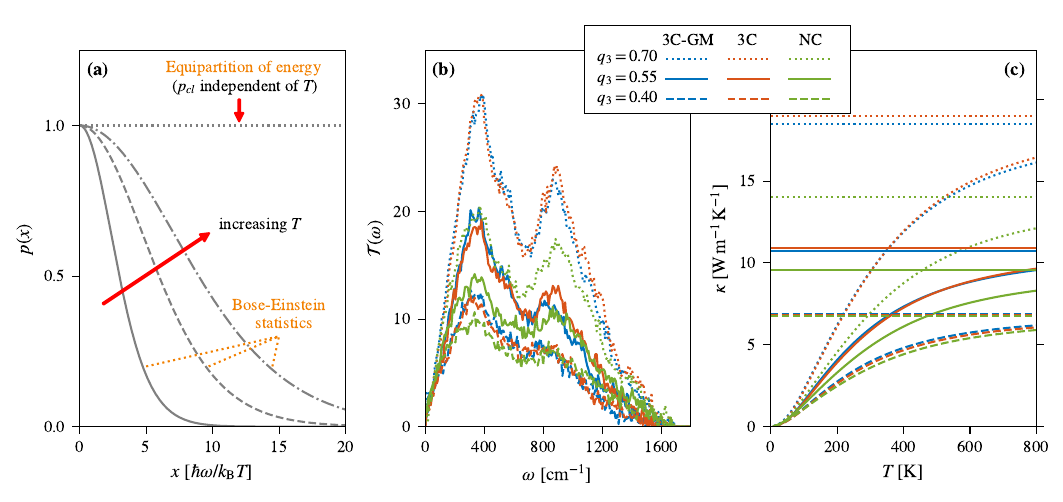}
	\caption{
		Classical and quantum mechanical weight functions are plotted as a function of $x=\hbar\omega/\kB T$ (a). The classical weight function is equal to unity independent of temperature, whereas the quantum mechanical one peaks at $x=0$ and approaches $p_{cl}$ at the high-temperature limit.
		Transmission (b) and conductivity (c) for different structure types and $q_3$ values. Blue, red, and green colors represent 3C-GM, 3C, and nC schemes, while dotted, solid, and dashed curves stand for $q_3$ values of 0.7, 0.55, and 0.4, respectively.
		In (b), the horizontal lines depict thermal conductivity values at the classical limit.
	}
	\label{fig:trans_cond}
\end{figure*}

In Fig.~\ref{fig:trans_cond}(b), we present the vibrational
transmission spectra of MAC for three representative
\(q_3\) values, namely 0.70, 0.55, and 0.40. Different
amorphization schemes are distinguished by color (blue for
3C-GM, red for 3C, and green for nC), whereas dotted, solid,
and dashed lines correspond to \(q_3 = 0.70\), 0.55, and 0.40,
respectively. For all schemes, the transmission decreases
across the entire spectrum as \(q_3\) is reduced, with
high-frequency modes being suppressed more strongly than
low-frequency ones. A comparison of the 3C-GM and 3C spectra
shows that the absence of three- and four-membered rings has
a negligible influence, as their transmission functions remain
nearly identical at each \(q_3\) value. By contrast, nC
configurations scatter intermediate-frequency vibrations more
effectively than 3C and 3C-GM structures, a distinction that
becomes increasingly pronounced at larger \(q_3\), owing to
the presence of voids and coordination defects in the nC
structures. In all cases, however, at fixed \(q_3\), the
transmission at both low and high frequencies remains
comparatively insensitive to the structural type.

Having analyzed the transmission spectra, we now turn to the
temperature dependence of the thermal conductivity.
Temperature-dependent thermal
conductivity for the different MAC types (3C, 3C-GM, and nC)
and selected \(q_3\) values is shown in
Fig.~\ref{fig:trans_cond}(c).

At low temperatures, thermal conductivity scales with $T^3$ in 3D crystalline solids. In 2D, however, linear acoustic modes' contribution scales as $T^2$, whereas that of the out-of-plane mode as $T^{3/2}$.~\cite{ziman2001electrons}
As a result, in 2D crystalline materials the exponent ranges between $1.5$ and $2$ depending on the relative contributions of in-plane and out-of-plane modes.

In many 3D amorphous materials, $\kappa\propto T^2$ behavior is commonly observed at low temperatures~\cite{zeller_prb_1971, anderson_philmag_1972}.

Within the harmonic Landauer framework used here, we find effective exponents ranging between 1.2 and 1.8 with larger $q_3$ values yielding larger exponents.
The appearance of a plateau in TC at low temperatures is a well-known phenomenon in 3D amorphous materials.
No clear plateau is observed in our computations of MAC. Instead, a smooth crossover connects the low- and high-temperature regimes, with \(\kappa(T)\) strongly
dependent on \(q_3\).
Unlike crystalline solids, amorphous materials do not exhibit a
pronounced decrease of thermal conductivity at high temperatures.
At elevated temperatures, \(\kappa(T)\) saturates to a constant value whose magnitude depends on the structural details and ranges from 6.7 to 18.9~\(\WmK\) for the structures considered here.

Comparing the temperature-dependent TC of different structural types,
for \(q_3 = 0.70\), the CRN structures (3C and 3C-GM) exhibit
nearly identical thermal conductivities over the entire
temperature range, whereas the NC@RN structure shows a
substantially lower \(\kappa\), consistent with the behavior
inferred from the transmission spectra. Nevertheless, NC@RN at
\(q_3 = 0.70\) still exhibits a higher thermal conductivity
than the CRN structures at \(q_3 = 0.55\). As \(q_3\)
decreases, the difference in \(\kappa(T)\) between CRN and
NC@RN structures diminishes, and for \(q_3 = 0.40\) their thermal
conductivities become nearly indistinguishable across the full
temperature range.
These results indicate that with increasing amorphization the influence of structural type, whether CRN or NC@RN, on thermal conductivity becomes progressively weaker. In contrast, \(q_3\) itself emerges as a robust indicator of thermal transport in highly amorphized MAC samples.

To establish direct contact with classical simulations and to access
the high-temperature limit, we additionally evaluate the conductivity
in the classical limit of the Landauer formalism.

\begin{figure}[b]
    \centering
    \includegraphics[width=0.75\textwidth]{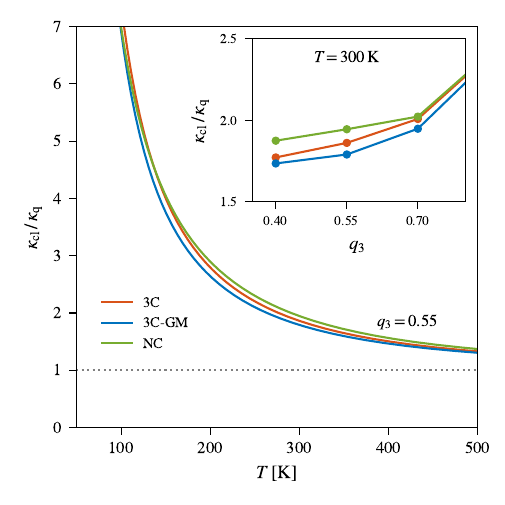}
    \caption{The ratio of classical and quantum thermal conductivities in MAC with $q_3=0.55$ is plotted as a function of temperature (main plot), and at room temperature for varying $q_3$ values (inset).  }
    \label{fig:K_oran}
\end{figure}

As shown in
Table~\ref{table:conductivity} and Fig.~\ref{fig:K_oran}, quantum
statistics reduces the room-temperature conductivity by a factor of
approximately 1.7--2.0, depending on \(q_3\).
{
It is also useful to compare the classical limit obtained here with previous classical molecular dynamics studies. Reported room-temperature TC values from classical MD vary widely: some studies predict values substantially higher than our classical-limit conductivities~\cite{bazrafshan-2017-ijhmt}, whereas others report values around 10–15~$\WmK$~\cite{mortazavi_carbon_2016,antidormi-2020-2dmat,zhang-2022-apl}, falling within the range obtained here. This spread likely reflects differences in structural models, amorphousness, sample size, boundary conditions, interatomic potentials, and the treatment of finite-size effects. Since the degree of amorphousness is not always quantified consistently across earlier simulations, a strict one-to-one comparison is not possible. Nevertheless, the overlap between part of the MD literature and our classical-limit values supports the physical range of the present calculations.
We also note that the room-temperature quantum-to-classical reduction obtained here for MAC is comparable to that reported by Lv and Henry for amorphous carbon films, where applying a quantum heat-capacity correction reduced the Green–Kubo thermal conductivity by approximately a factor of three~\cite{lv-2016-apl}.
Overall, this comparison shows that structural variability already produces a broad range of classical TC values, while quantum statistics further renormalize these values downward at room temperature.
}

\section{Conclusion}
\label{sec:conclusion}

In conclusion, we have shown that heat conduction in two-dimensional monolayer amorphous carbon is controlled by the combined effects of structural topology, degree of amorphization, and quantum statistics. By comparing continuous random networks and nanocrystallite-embedded random networks over a range of $q_3$ values, and by evaluating the corresponding classical limit within the same Landauer framework, we identify how structural disorder shapes the transmission spectrum while quantum statistics renormalize the absolute thermal conductivity. Within the harmonic framework and temperature range considered here, we observe an effective low-temperature behavior of $T^\alpha$ with $\alpha$ ranging between 1.2 and 1.8 depending on the amorphousness of the system.
At high temperatures, TC is nearly temperature independent, as expected for amorphous systems.~\cite{zhu-2016-nanoletters}
For relatively more ordered systems (higher order parameter, $q_3$), continuous random networks maintain higher thermal conductivities compared to nanocrystallites embedded in random networks.
For highly amorphous samples (lower $q_3$ values), the detailed structural motif becomes secondary, indicating that, in the strong-disorder limit, thermal transport approaches a unified behavior governed primarily by the global degree of amorphization rather than by microscopic structural details.
Although a direct one-to-one comparison with experiment is not yet possible because the degree of amorphousness was not quantified in the available measurements, the reported experimental thermal conductivities fall within the range predicted here.
The present approach is free from the intrinsic limitations of classical simulations and provides a rigorous and internally consistent description of vibrational heat transport.
Moreover, the classical limit is recovered directly within the Landauer framework, enabling a consistent comparison between quantum and classical thermal conductivities.
Together, these results position MAC as a useful two-dimensional amorphous platform for studying the interplay between topology-driven vibrational scattering and quantum-statistical heat transport.

\clearpage

\begin{figure*}
	\includegraphics[width=\textwidth]{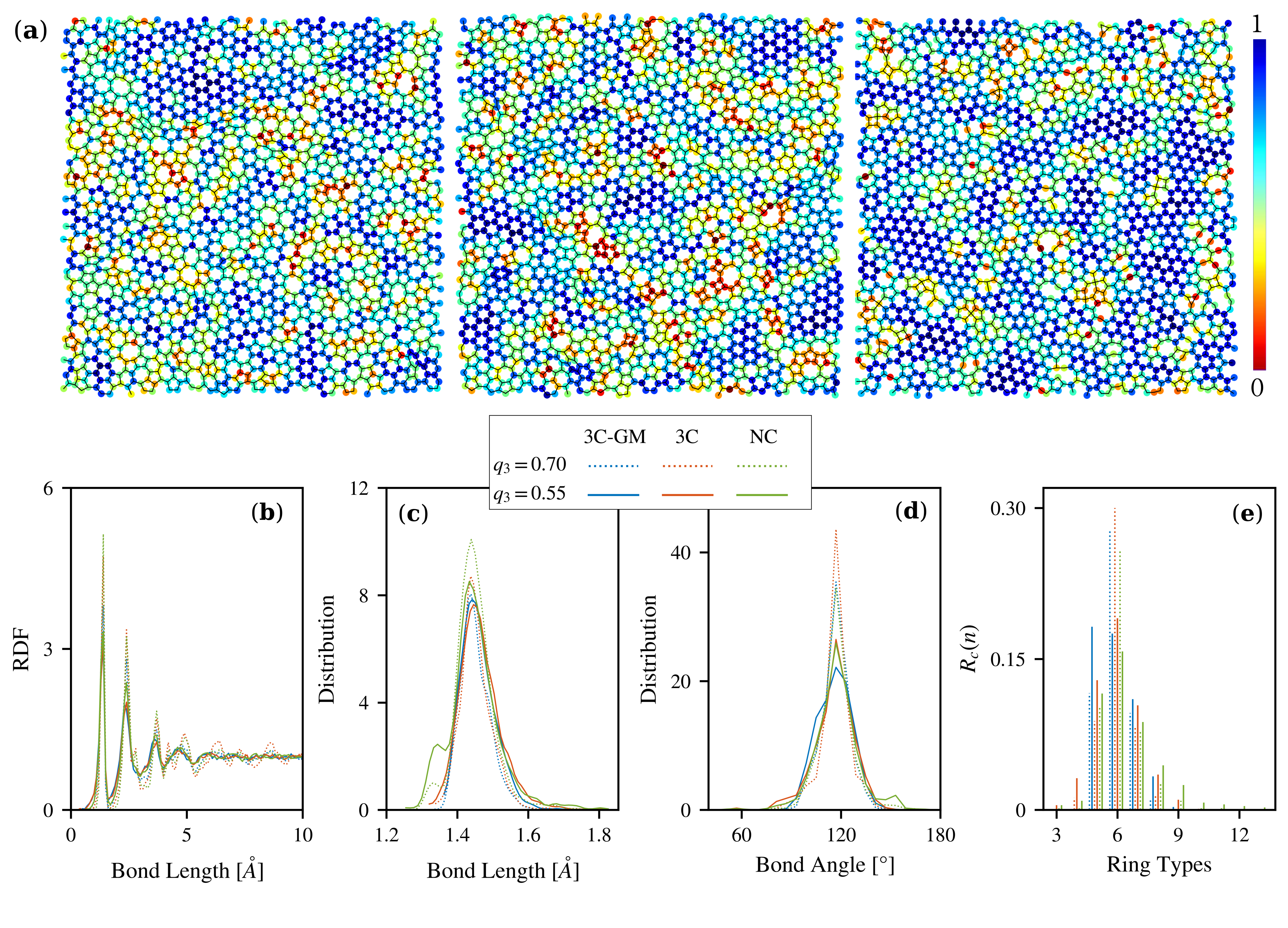}
	\caption{(a) Continuous random network (CRN) and embedded nanocrystallite (NC@RN) structures obtained using 3C-GM, 3C, and nC algorithms are shown in the left, center, and right panels. The colors represent the value of the local bond order parameter, $q_3$.
    The average $q_3$ value is 0.6 and it is the same for all structures shown in (a).
    Radial distribution function (b), bond length distribution (c), bond angle distribution (d), and ring statistics (e) of the configurations with different structural types and two different $q_3$ parameters are shown in the lower panels.
	}
	\label{fig:structural}
\end{figure*}

\clearpage

\clearpage

\clearpage

\clearpage

\clearpage

\begin{table}[h]
	\begin{tabular}{p{0.15\textwidth}p{0.35\textwidth}}
		Symbol & Explanation  \\
		\hline
		$T$ & temperature  \\
		$\mathcal{T}(\omega,T)$ & transmission amplitude  \\
		$G(T)$ & thermal conductance \\
		$p(\omega,T)$ & quantum mechanical weight function \\
		$p_{cl}$ & classical weight function \\
		$\kappa(T)$ & thermal conductivity \\
		$\kappa_{cl}$ & classical thermal conductivity \\
		$f_\mathrm{BE}(\omega,T)$ & Bose-Einstein distribution  \\
		$k_B$ & Boltzmann constant\\
		$L$ & length of the amorphous region \\
		$A$ & cross section area of the amorphous region\\
	\end{tabular}
\end{table}

\begin{table}[h]
	\begin{tabular}{p{0.15\textwidth}p{0.35\textwidth}}
		Abbreviation & Explanation  \\
		\hline
		2D & two-dimensional  \\
		3D & three-dimensional  \\
		MAC & monolayer amorphous carbon \\
		TC & thermal conductivity, $\kappa$ \\
		AIREBO & adaptive intermolecular reactive empirical bond order \\
		EDKMC & energy-driven kinetic Monte Carlo \\
		C2A   & crystalline to amorphous\\
		R2C & random to crystalline \\
		3C  & 3-coordinated \\
		3C-GM & 3-coordinated geometric-model \\
		nC & non-constrained\\
		CRN & continuous random network \\

        NC@RN & Nanocrystallite inside random network\\
		RDF & radial distribution  function\\
        PR & Participation ratio \\
		IPR & Inverse participation ratio
	\end{tabular}
\end{table}

\clearpage

\clearpage\clearpage
\onecolumngrid
\setcounter{equation}{0}
\setcounter{figure}{0}
\setcounter{section}{0}
\setcounter{table}{0}
\setcounter{page}{1}

\renewcommand\theequation{S\arabic{equation}}
\renewcommand\thefigure{S\arabic{figure}}
\renewcommand\thetable{S\Roman{table}}
\renewcommand\thepage{S\arabic{page}}
\renewcommand\thesection{S-\Roman{section}}

\nolinenumbers

\begin{center}
	\fontsize{16}{30} \selectfont\sffamily{\supplement}\\
	\fontsize{20}{30} \selectfont\sffamily\textbf{\baslik}
\end{center}

\setcounter{linenumber}{1}

\section{\supplement\ On Structure Generation And Characterization}

We employ a combination of the AIREBO (Adaptive Intermolecular Reactive Empirical Bond Order) and the re-optimized Tersoff potentials for modeling the structures and obtaining the force constant matrices.
The AIREBO potential, parametrized for hydrocarbons, is used during structural optimization.
The simulation cell is subsequently rescaled, and all atoms are fully relaxed using the re-optimized Tersoff potential. Final structural energies and interatomic force constants are obtained with the re-opt Tersoff potential. Relaxations are carried out using the conjugate-gradient algorithm as implemented in the LAMMPS package,\cite{LAMMPS} with full cell relaxation allowed until the residual force on each atom is below 10$^{-3}$~ eV/$\Ang$.

Amorphous structures are generated using two different algorithms, whose initial configurations and intermediate steps follow opposite approaches.
The first approach, namely the crystalline-to-amorphous (C2A) approach, starts with a perfect crystalline structure and applies bond rotations, provided that the total energy increases after each step.
The second starts with a random distribution of atoms and applies bond rotations that lower the energy, hence it follows a random to crystalline (R2C) path.
Details about the implementation of these approaches are given below.

\begin{table*}[b]
	\centering
	\begin{ruledtabular}
		\begin{tabular}{l   c c c}
			\multirow{3}{*}{}        &
			\multicolumn{3}{c}{Structural Type}\\

             	& 3C-GM   & 3C    & nC \\
			\cmidrule{2-4}
            Constraints & 3-coordinated (general model) & 3-coordinated & non-constrained
            \\

			Procedure &     Crystalline to amorphous   & Crystalline to amorphous    & Random to ordered \\

			Character  &  CRN             & CRN & NC@CRN \\

			Coordination Defects & \xmark & \xmark & \cmark \\

			Rings With Less Than Five Members & \xmark & \cmark & \cmark\\

			Rings With More Than Ten Members & \xmark & \xmark & \cmark\\

			Voids & \xmark & \xmark & \cmark\\
		\end{tabular}
	\end{ruledtabular}
    \caption{Structural classes and their characteristics.}
    \label{table:structuraltypes}
\end{table*}

\subsection{Crystalline-to-Amorphous (C2A) Schemes: Continuous Random Networks}

During the C2A procedure, new configurations that lowered the energy ($E_f < E_i$) are rejected.
Otherwise, acceptance is decided with a Metropolis criterion.
The threshold energy is chosen such that approximately half of the attempted configurations are accepted.
In addition, each atom is required to be strictly three-coordinated within a bond radius less than 2.0~Å.
Two variants of this procedure were implemented. In the 3C-GM version, stricter geometric constraints were imposed by reducing the cutoff to 1.98~Å, yielding networks structurally similar to geometric modeling (GM) approaches.~\cite{kumar-2012-jopcm}
These networks are composed primarily of five-, six-, and eight-membered rings.
In contrast, the 3C variant permitted smaller three- and four-membered rings while still enforcing threefold coordination, thereby producing a slightly more disordered network.

\begin{figure}[t]
	\vspace{3em}
	\centering
	\includegraphics[width=0.75\textwidth]{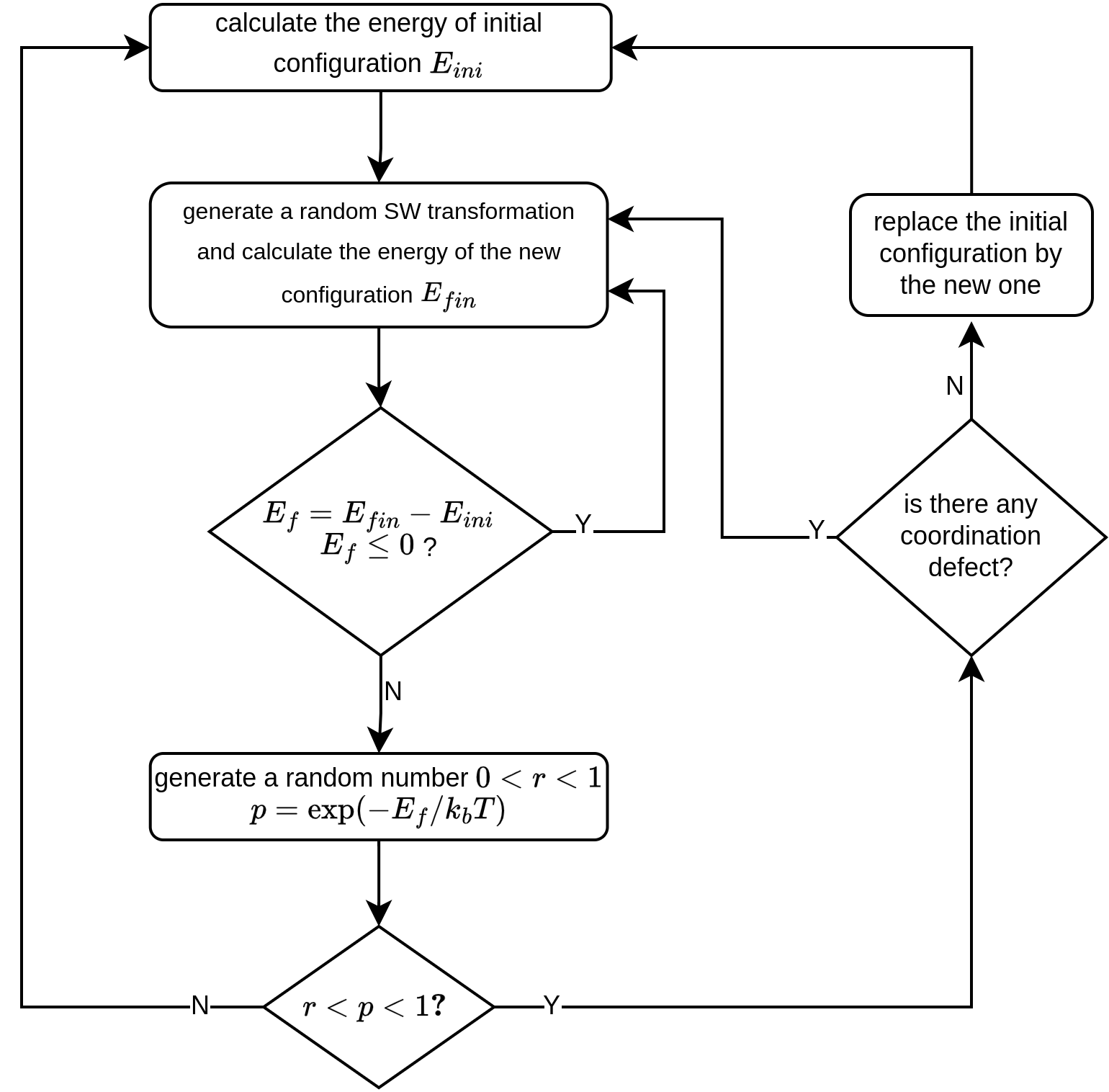}
	\caption{The flowchart of the amorphization algorithm for continuous random network structures.}
	\label{fig:MCflowchart}
\end{figure}

\subsection{Random-to-Crystalline (R2C) Scheme: Nano-crystallite in Continuous Random Network}

The R2C structures are generated using Zhuang’s crystallization scheme, which employs Energy-Driven Kinetic Monte Carlo (EDKMC) algorithm~\cite{zhuang-2016-pccp,toh_nature_2020}

in which any bond rotation that lowered the total energy was automatically accepted. Otherwise, acceptance was determined by the Metropolis criterion at $T=6000$~K, comparing a probability factor $\rho = \exp[(E_i - E_f)/k_B T]$ with a randomly generated number\cite{metropolis-1953-tjocp}. The R2C scheme produced networks with distinctly different features compared to the 3C and 3C-GM schemes. The resulting structures are not Zachariasen-type glasses but  contain coordination defects, voids, three- and four-membered rings, and nano-crystallite islands embedded within a continuous random network.

To apply the Green's function (GF) method, structures must be split into three regions: the central device, the left, and the right reservoirs; see Fig.~\ref{sfig:GFpartition}. Reservoirs are two semi-infinite crystals, and the central region, in our case, a finite disordered region, determines the system's thermal behavior. Both algorithms were applied to the device region of the structures to apply the Green's function method.

\newpage

\subsection{Local Bond Order Parameter (\texorpdfstring{\boldmath$q_3$\unboldmath}{q_3})}

The structures in Fig.~\ref{fig:structural} are color-coded using a color map that reflects the local bond order parameter associated with individual carbon atoms, which is calculated by
\begin{align}
	\Tilde{q}_l(i) &= \frac{1}{N_i}
	\sum\limits_{j\epsilon \mathcal{S}_i}
	\boldsymbol{q}_l(i) \cdot \boldsymbol{q}_l(j),\\
	\Tilde{q}_3(i) &=  \frac{1}{N_i}  \sum\limits_{j\epsilon\mathcal{S}_i}
	\sum\limits_{m=-3}^{3}
	\frac{q_{3m}(i)}
	{\sqrt{\sum\limits_{m=-3}^{3}\left|q_{3 m}(i)\right|^2}}
		\cdot
	\frac{q_{3m}^*(j)}
	{\sqrt{\sum\limits_{ m= -3}^{3}\left|q_{3 m}(j)\right|^2}},
\end{align}
where the first summations are over the first nearest neighbors.
This parameter measures the degree of local structural ordering, which is practical for measuring the  degree of amorphousness.
In the color map, $q_3=1$ is represented in blue, indicating an ideal local hexagonal arrangement. In contrast, $q_3=0$ is shown in red, indicating a complete loss of local hexagonal symmetry.

We compute the local bond order parameters for structures analyzed using the Green's function method by averaging only over atoms within the central scattering region. Atoms belonging to the adjacent buffer and reservoir regions are excluded from this averaging to ensure that the calculated parameters reflect only the disorder within the device region of the system.
Atoms that belong to crystalline parts of buffers may exhibit $q_3$ values slightly below one. This  is mainly due to out-of-plane distortions that locally disrupt the ideal planar geometry.

\begin{figure}[b]
	\centering
	\includegraphics[width=0.5\linewidth]{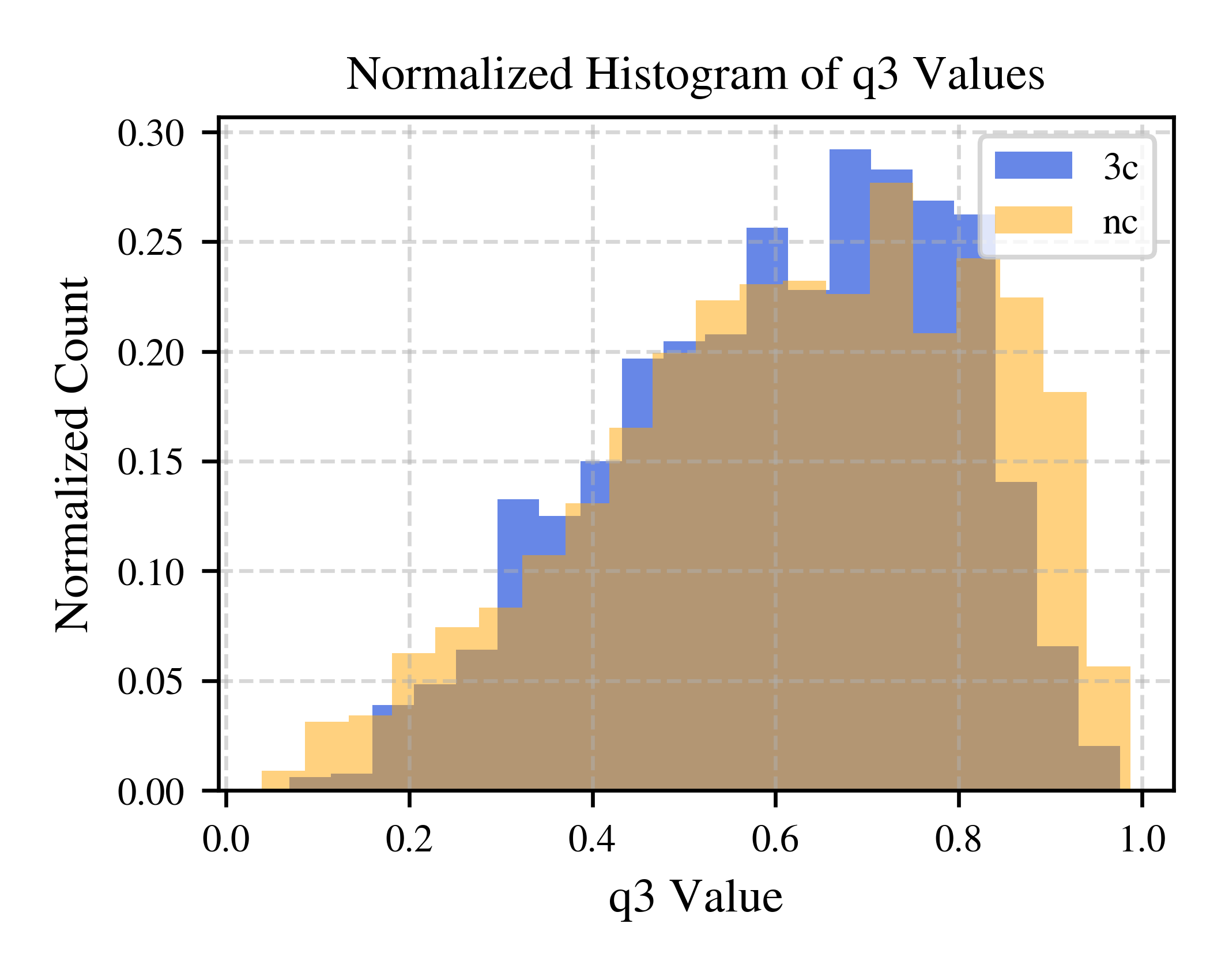}
	\caption{\label{fig:frog}Histograms of the atomic $q_3$ values in samples with $q_3$=0.55. }
\end{figure}

The histograms of the atomic $q_3$ values for different types of configurations are plotted in Fig.~\ref{fig:frog} to illustrate how disorder is distributed within the configurations. The sample sizes are $25 \times 25 \; \text{nm}^2$, with $q_3 = 0.55$. The 3C (nC) type configuration has 23,200 (24,700) atoms.
NC@RN configurations are more populated than CRNs at both low and high $q_3$ values.
In contrast,  $q_3$ of CRNs are mainly populated within the range [0.25, 0.85], which reflects its continuous random structure without voids and coordination defects.

\clearpage
\subsection{Images of CRN and NC@RN MAC Structures with Various \texorpdfstring{\boldmath$q_3$\unboldmath}{q_3} Values}
Below we show a selection of 3C, 3C-GM and NC@RN type structures with various $q_3$ values. We note that these are the actual structures that are used in quantum transport calculations, hence include buffer layers, pristine layers that belong to the reservoirs, as well as the amorphous scattering regions.

\begin{figure}[b]
	\centering
	\begin{tabular}{@{}c@{}}
		\includegraphics[height= 0.22\textheight]{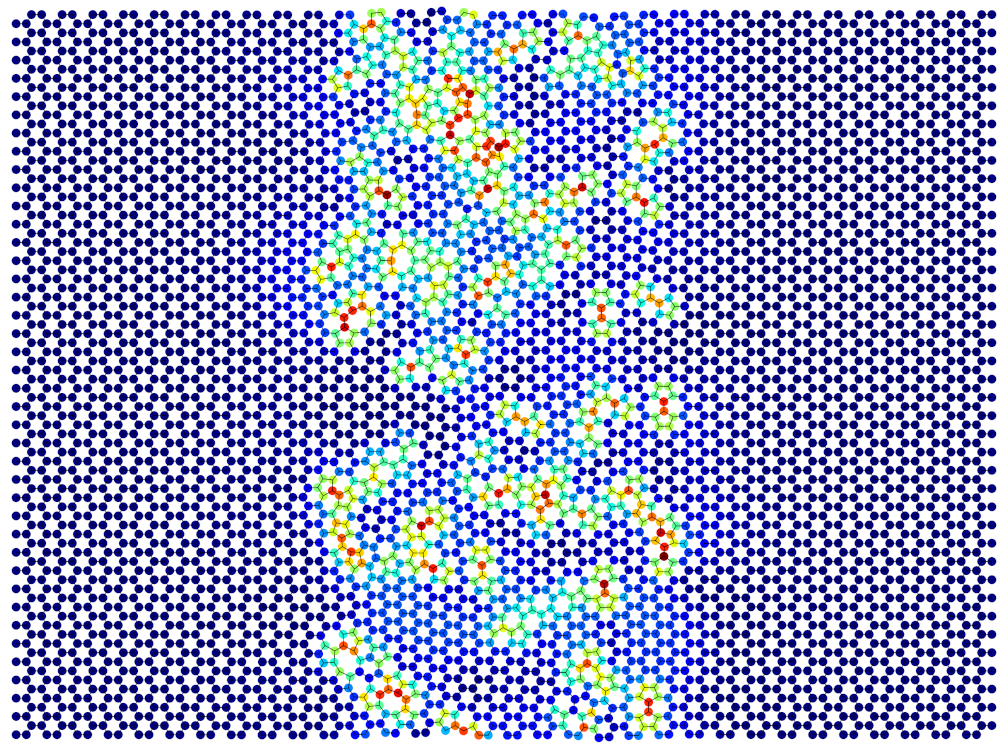} \\[\abovecaptionskip]
		\small (a) $q_3=$ 0.70, 149th step
	\end{tabular}

	\vspace{\floatsep}

	\begin{tabular}{@{}c@{}}
		\includegraphics[height = 0.22\textheight]{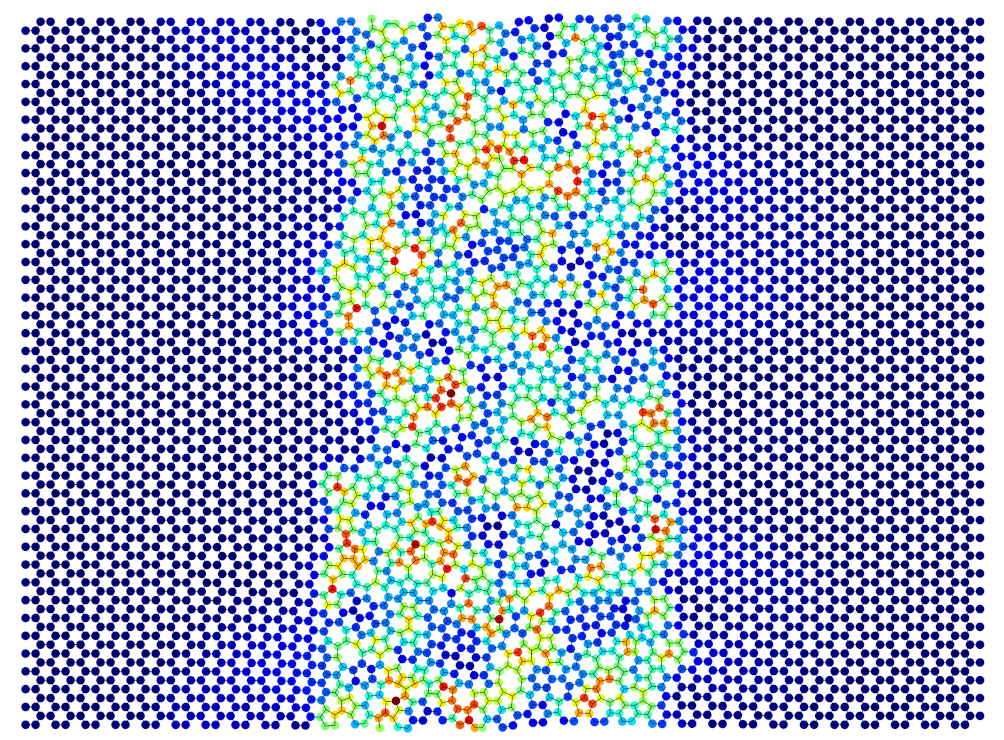} \\[\abovecaptionskip]
		\small (b) $q_3=$ 0.55, 553th step
	\end{tabular}
	\vspace{\floatsep}

	\begin{tabular}{@{}c@{}}
		\includegraphics[height = 0.22\textheight]{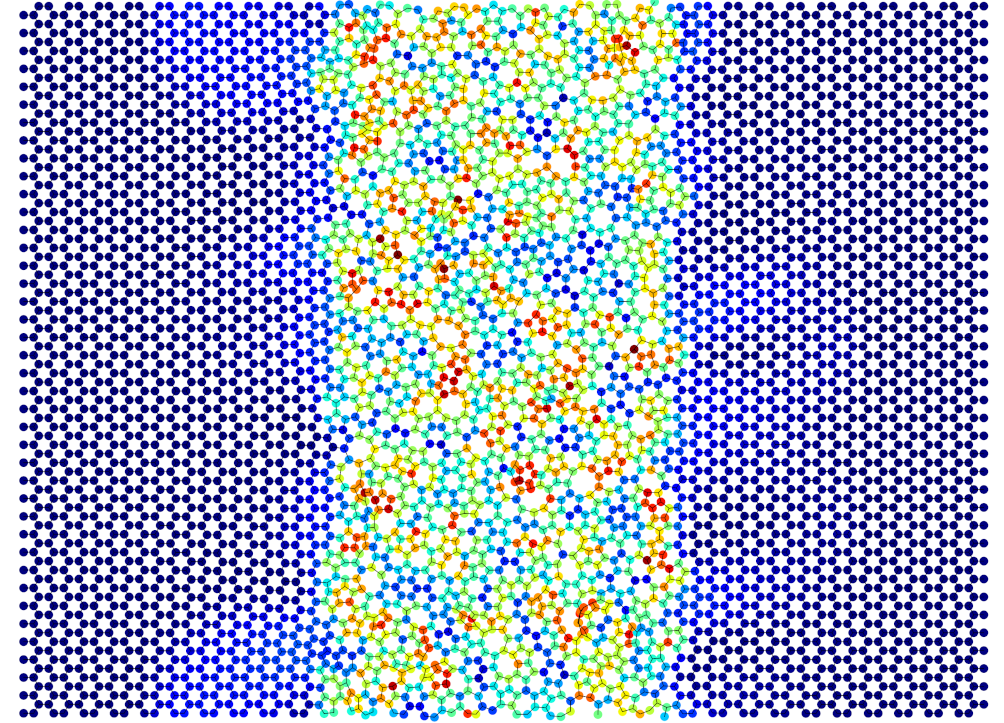} \\[\abovecaptionskip]
		\small (c) $q_3=$ 0.40, 8441th step
	\end{tabular}

	\caption{3C-GM type configurations}\label{sfig:3cgmconfigs}
\end{figure}

\begin{figure}
	\centering
	\begin{tabular}{@{}c@{}}
		\includegraphics[height= 0.25\textheight]{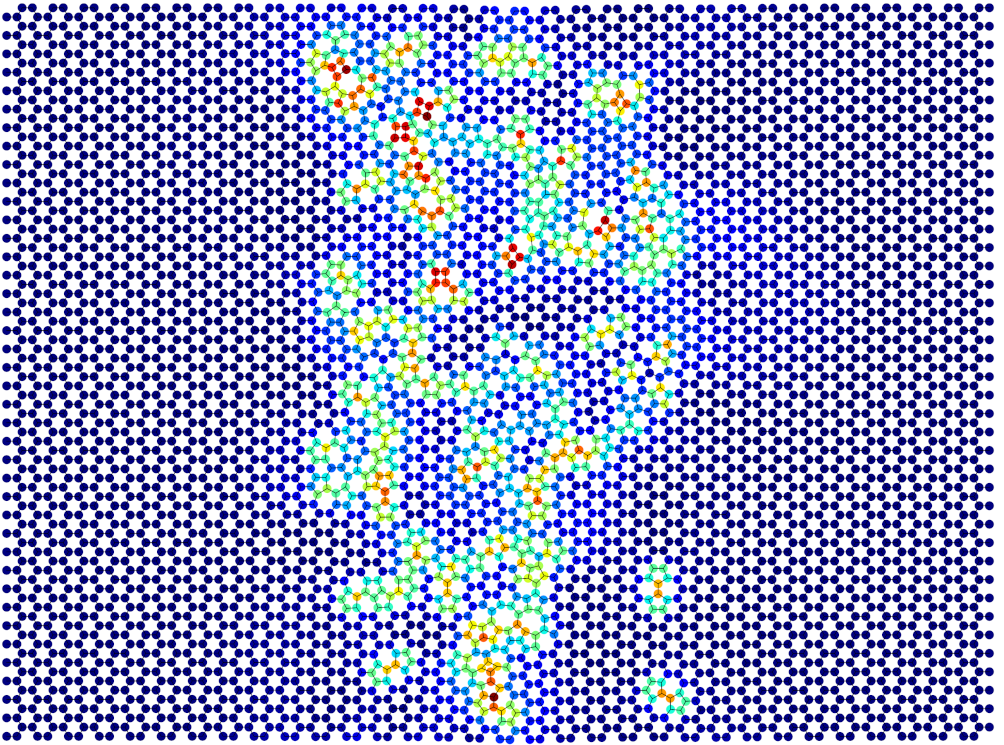} \\[\abovecaptionskip]
		\small (a)$q_3=$ 0.70, 126th step
	\end{tabular}

	\vspace{\floatsep}

	\begin{tabular}{@{}c@{}}
		\includegraphics[height = 0.25\textheight]{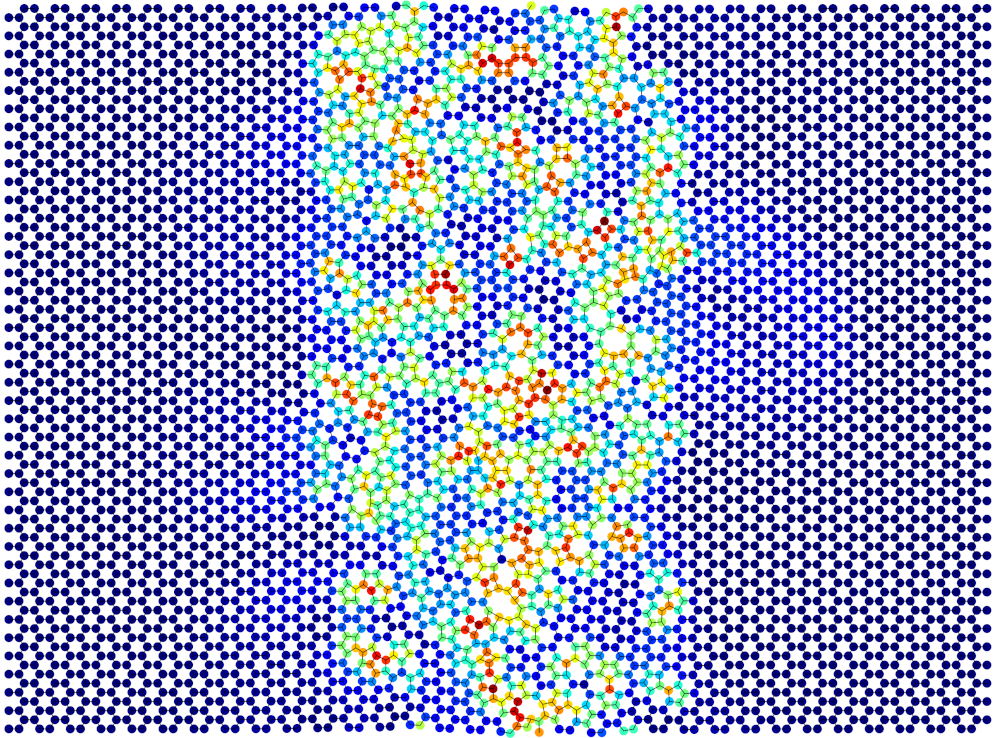} \\[\abovecaptionskip]
		\small (b) $q_3=$ 0.55, 249th step
	\end{tabular}
	\vspace{\floatsep}

	\begin{tabular}{@{}c@{}}
		\includegraphics[height = 0.25\textheight]{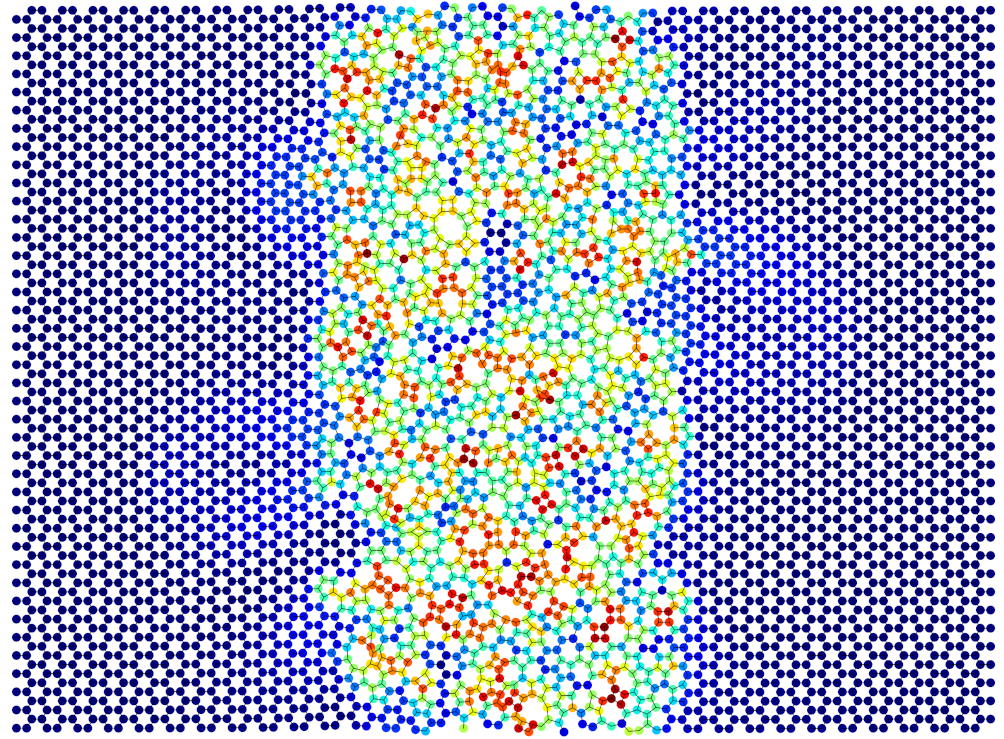} \\[\abovecaptionskip]
		\small (c) $q_3=$ 0.40, 500th step
	\end{tabular}

	\caption{3C type configurations}\label{sfig:3cconfigs}
\end{figure}
\begin{figure}
	\centering
	\begin{tabular}{@{}c@{}}
		\includegraphics[height= 0.25\textheight]{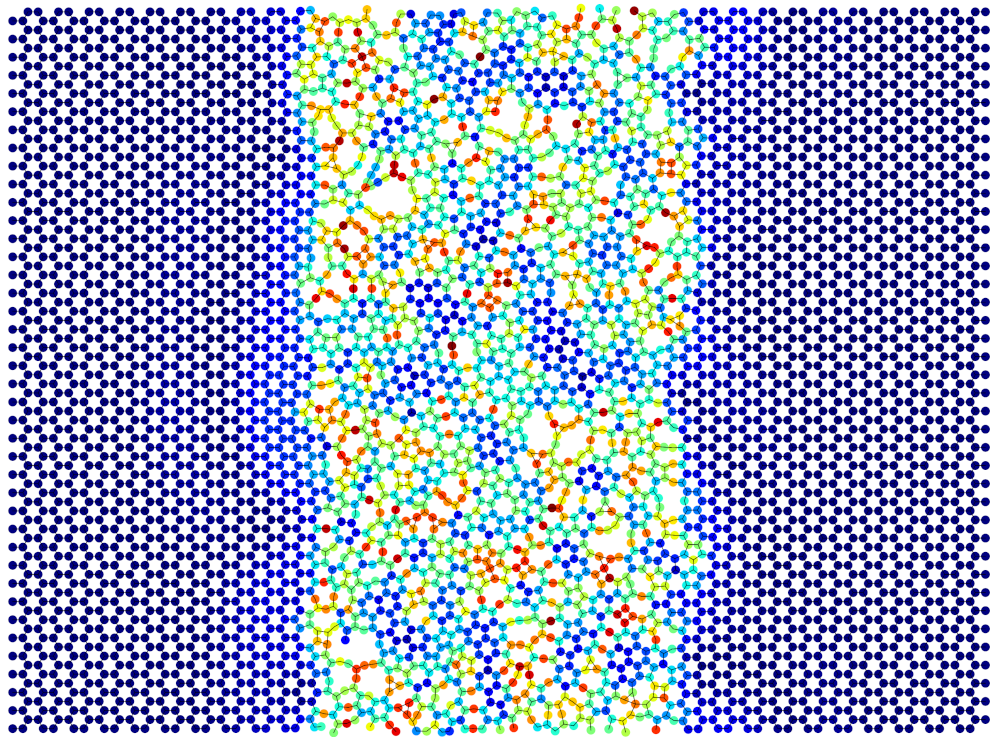} \\[\abovecaptionskip]
		\small (a) $q_3=$ 0.40, 5362th step
	\end{tabular}

	\vspace{\floatsep}

	\begin{tabular}{@{}c@{}}
		\includegraphics[height = 0.25\textheight]{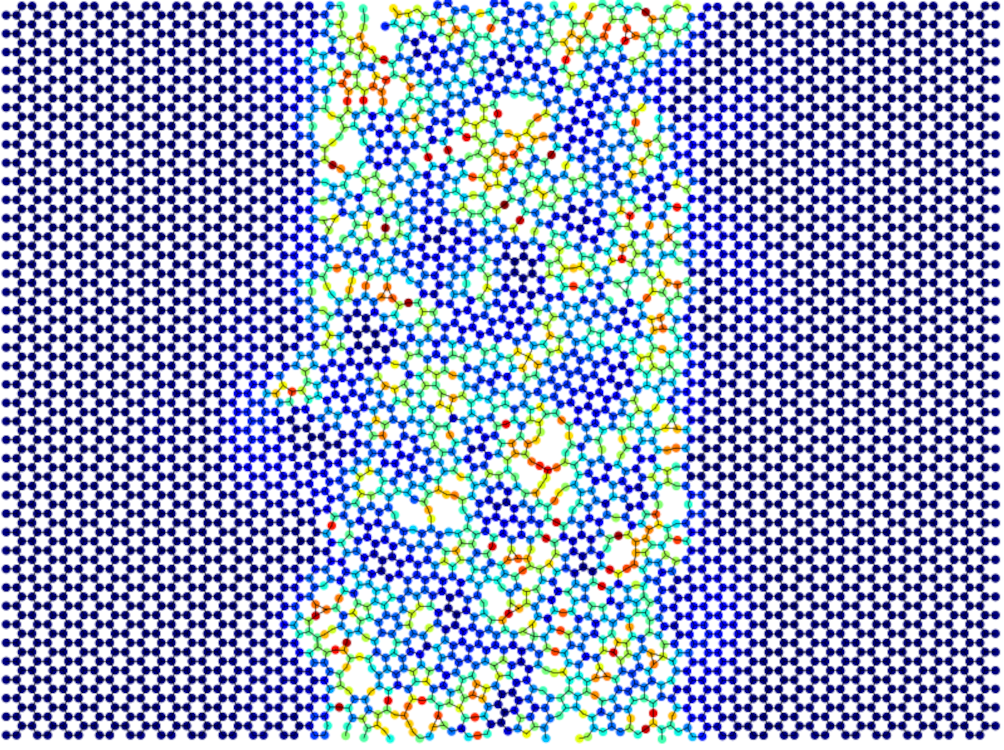} \\[\abovecaptionskip]
		\small (b) $q_3=$ 0.55, 26152th step
	\end{tabular}
	\vspace{\floatsep}

	\begin{tabular}{@{}c@{}}
		\includegraphics[height = 0.25\textheight]{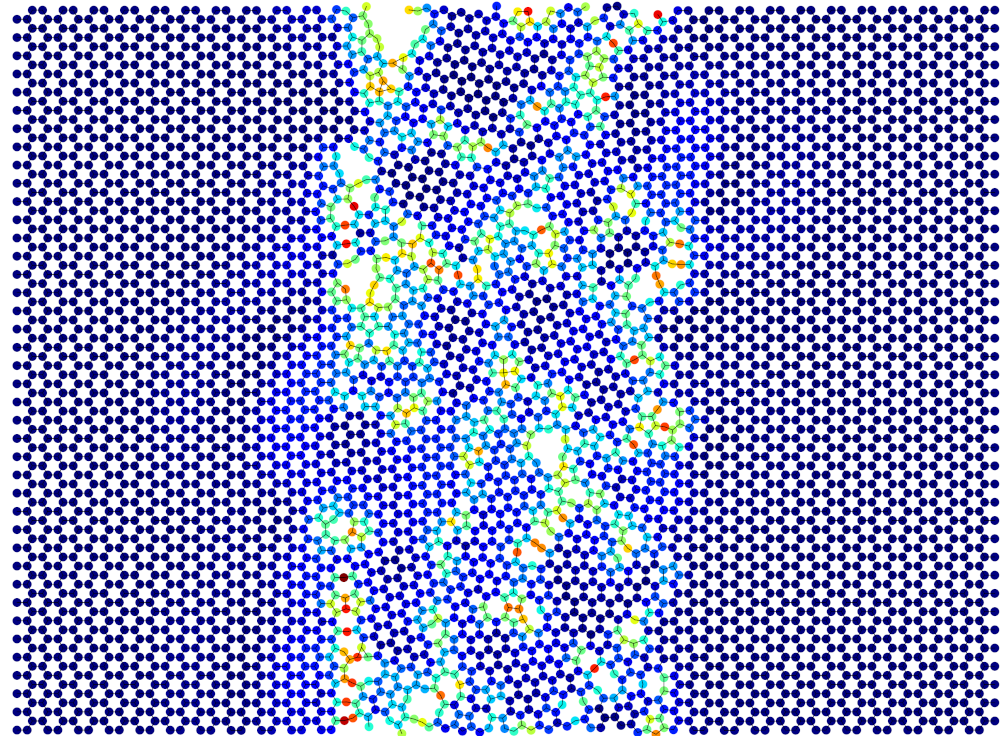} \\[\abovecaptionskip]
		\small (c) $q_3=$ 0.70, 93996th step
	\end{tabular}

	\caption{nC type configurations.}\label{sfig:ncconfigs}
\end{figure}

\begin{figure}
	\centering
	\begin{tabular}{@{}c@{}}
		\includegraphics[height= 0.25\textheight]{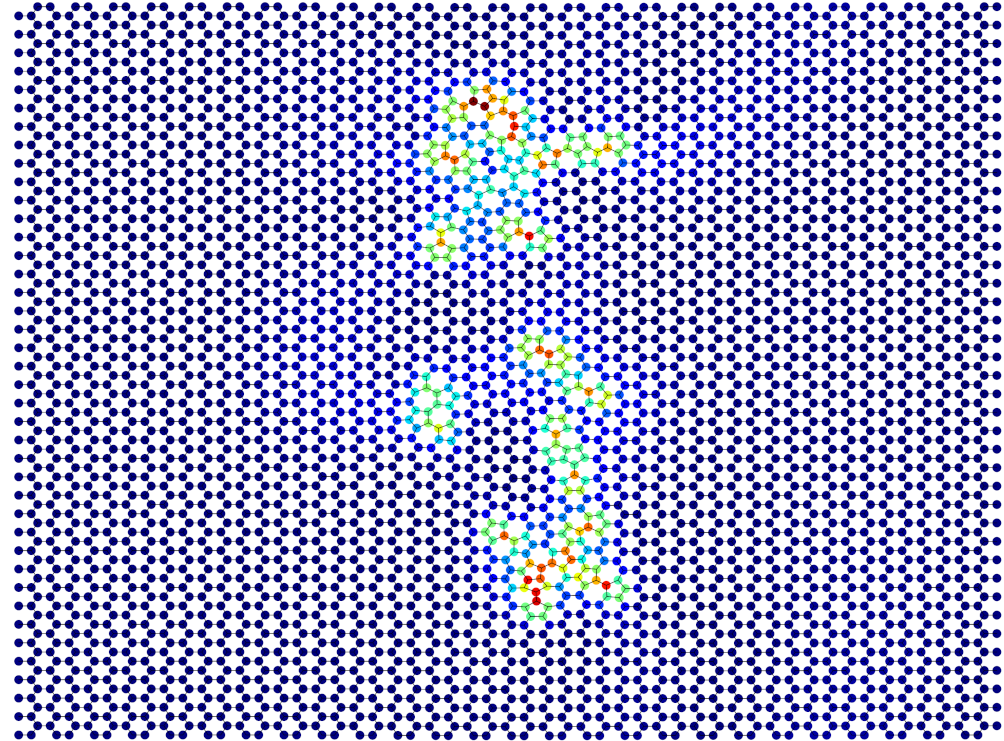} \\[\abovecaptionskip]
		\small (a) $q_3=$ 0.70, 36th step
	\end{tabular}

	\vspace{\floatsep}

	\begin{tabular}{@{}c@{}}
		\includegraphics[height = 0.25\textheight]{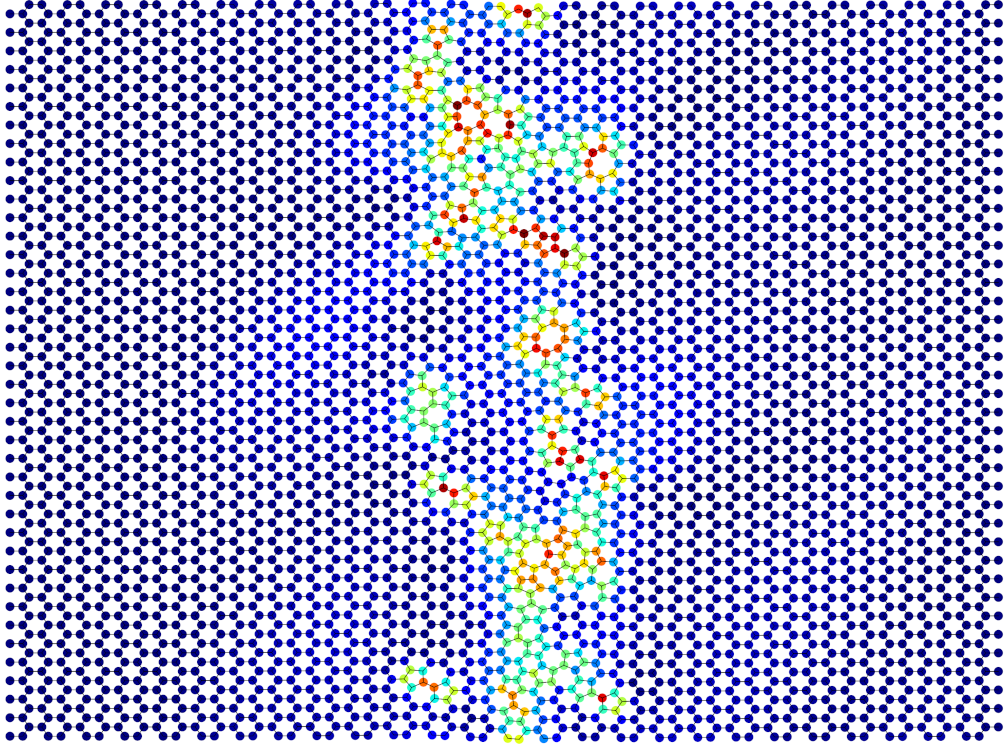} \\[\abovecaptionskip]
		\small (b) $q_3=$ 0.55, 64th step
	\end{tabular}

	\caption{3C-GM configuration type, device length: 1.6 nm}\label{sfig:short}
\end{figure}

\begin{figure}
	\centering
	\begin{tabular}{@{}c@{}}
		\includegraphics[height= 0.25\textheight]{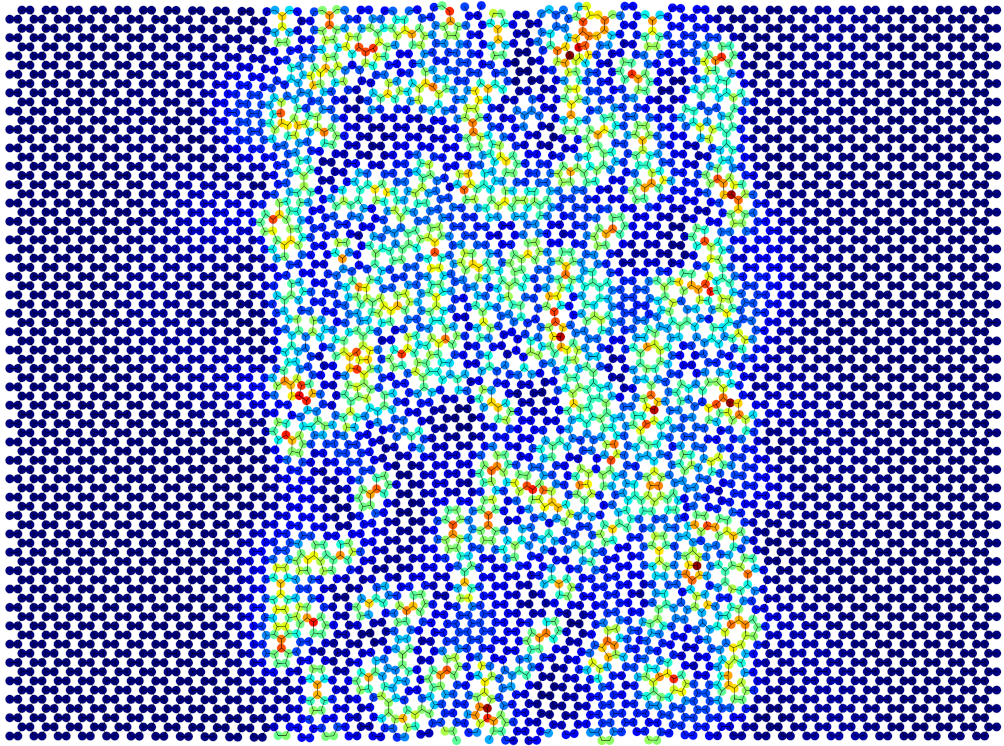} \\[\abovecaptionskip]
		\small (a) $q_3=$ 0.70, 357th step
	\end{tabular}

	\vspace{\floatsep}

	\begin{tabular}{@{}c@{}}
		\includegraphics[height = 0.25\textheight]{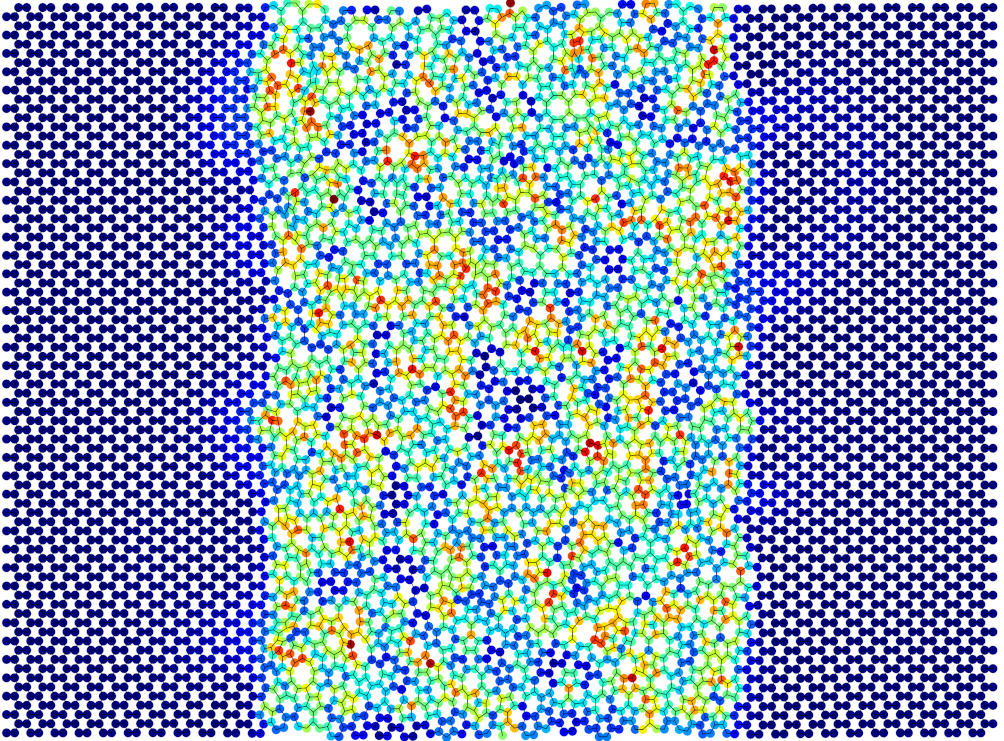} \\[\abovecaptionskip]
		\small (b) $q_3=$ 0.55, 8122th step
	\end{tabular}

	\caption{3C-GM configuration type, device length: 7.5 nm}\label{sfig:long}
\end{figure}

\clearpage
\subsection{Buckling Analysis}
Defects disturb the planarity of graphene and exhibit intrinsic ripples. The buckling analysis reveals the out-of-plane deviation from the planar structure of pure graphene. Here, the reservoirs' alignments are set to zero of the $z$-axis, and the color map's scale is in Angstroms.

\begin{figure}[h!]
	\includegraphics[height=75mm]{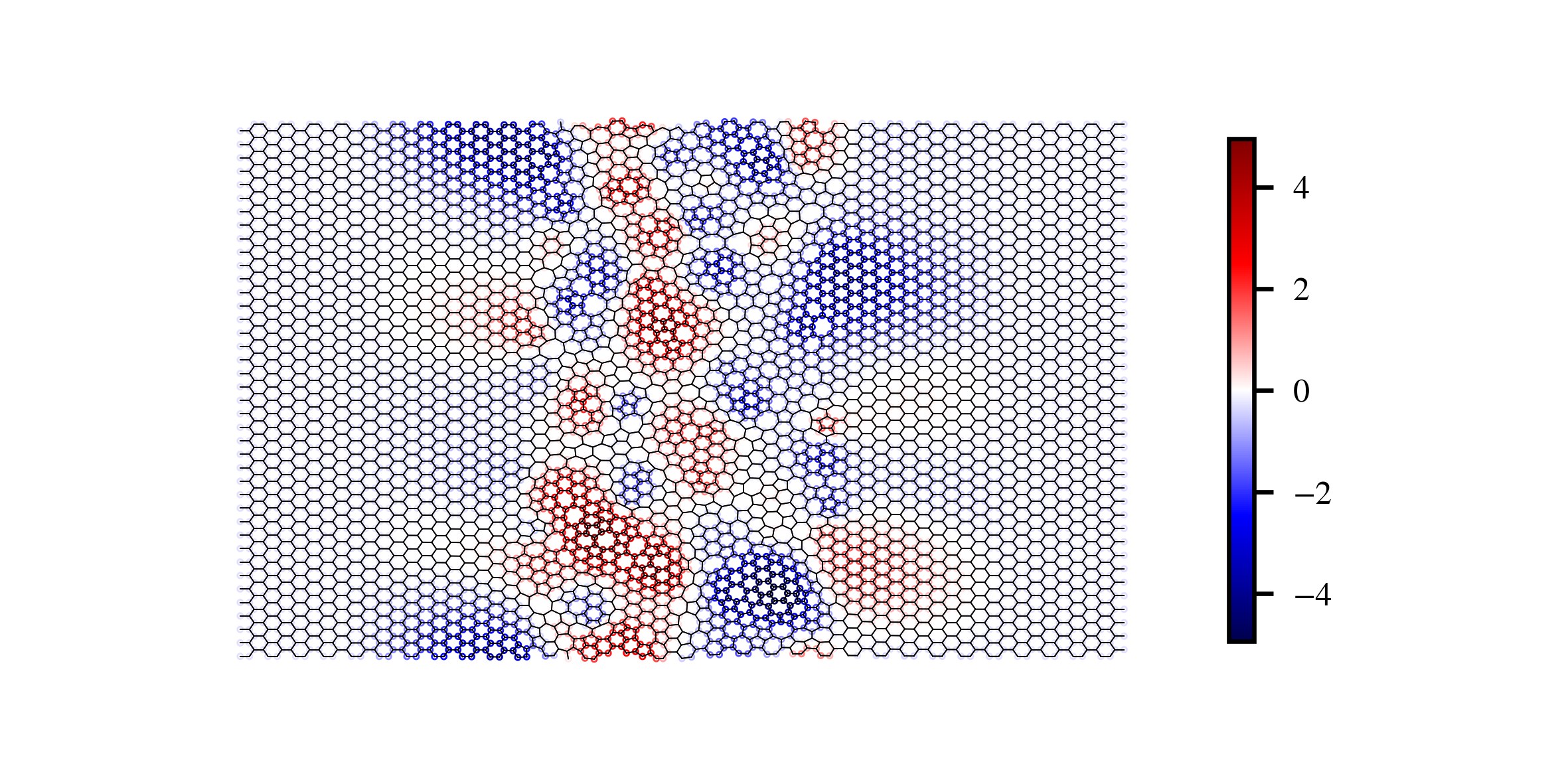}
	\caption{Buckling of 3C-GM structure with $q_3=0.55$. The colors indicate the displacement of atoms from the pristine graphene planes that make up the reservoirs.}
    \label{sfig:buckling}
\end{figure}

\begin{figure}[b]
	\includegraphics[width=100mm]{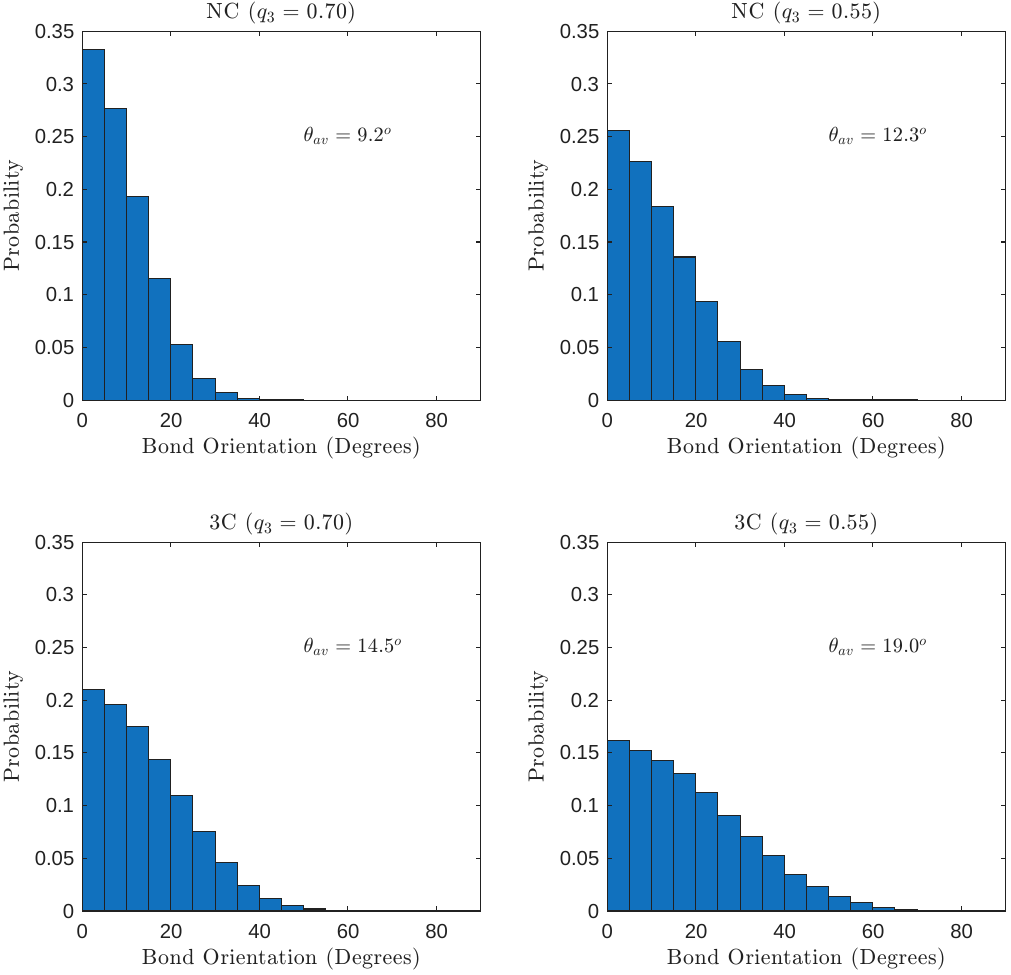}
	\caption{Bond orientation distributions of NC@RN and 3C-GM structures with $q_3=0.70$ and $q_3=0.55$.
	}\label{sfig:bucling_bondorientation}
\end{figure}

To quantitatively characterize the buckling behavior, we performed a bond orientation analysis on the two-dimensional structures. This method provides direct insight into the out-of-plane distortions of the system. In pristine crystalline graphene, all bonds lie strictly within the basal plane. In contrast, amorphous structures exhibit buckling, which results in finite angular deviations of the bonds from the plane.
The statistical distribution of bond angles with respect to the reference plane serves as a measure of the structural corrugation. Fig.~\ref{sfig:bucling_bondorientation} shows the bond orientation distributions of NC@RN and 3C-GM structures with lateral dimensions of $25 \times 25~\mathrm{nm}^2$ under periodic boundary conditions. For NC@RN structures with $q_3 = 0.70$, the mean bond angle relative to the plane is $9.2^\circ$, while this value increases to $12.3^\circ$ for $q_3 = 0.55$.
Applying the same analysis to 3C-GM structures yields larger average angles of $14.5^\circ$ and $19.0^\circ$, respectively. In NC@RN systems, the absence of topological constraints gives rise to a broader distribution of coordination numbers and ring sizes; nevertheless, the bond orientations remain closer to the plane, indicating a reduced degree of buckling compared to 3C-GM. This behavior can be attributed in part to the presence of crystallite regions in NC@RN structures, which locally suppress out-of-plane deformations.

\clearpage
\section{Interatomic Potentials and Vibrational Spectrum}

\begin{figure}[b]
	\centering
	\includegraphics[width=0.6\textwidth]
	{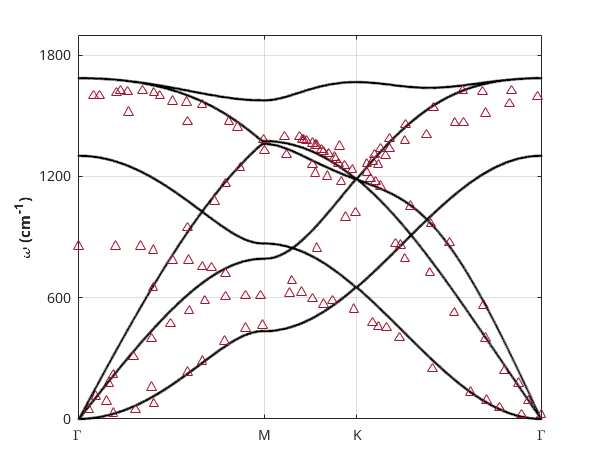}
	\includegraphics[width =0.6\textwidth]{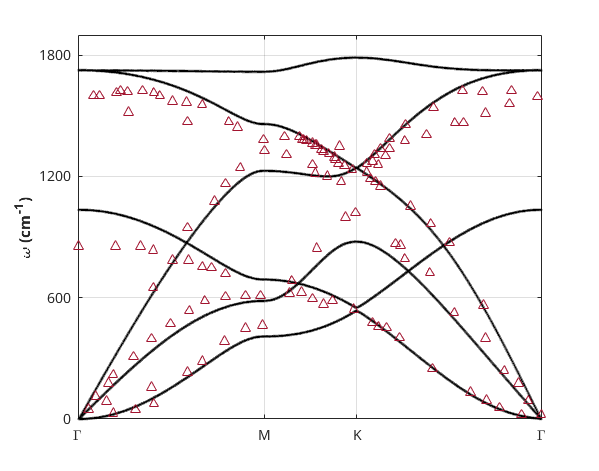}
	\caption{Phonon Dispersion relations. Triangles represent the inelastic X-ray scattering results for graphite.~\cite{mohr-prb-2007}. The phonon dispersion relation of graphene was determined with optimized Tersoff potential (top panel) and optimized Rebo-CH potential (bottom panel).}
	\label{sfig:pdrs}
\end{figure}

Phonon dispersions of graphene as obtained from optimized Tersoff~ \cite{tersoff_prb_1988,tersoff_prl_1988,lindsay-2010-prb} and Rebo-CH potentials are tested against available data to confirm the  reliability of interatomic potentials.
Periodic boundary conditions are imposed in all directions.
Distance between adjacent graphene layers is set to 50~\AA.
The structures are relaxed through the conjugate gradient algorithm with a force tolerance $10^{-3}$~eV/\AA.
The force constants are obtained within the finite difference method using LAMMPS software. The comparison of the obtained phonon dispersions is shown in Fig.~\ref{sfig:pdrs}.
Although phonon dispersions as obtained from Opt-Tersoff potential are in good agreement with the experimental data, the potential is not as accurate during the  crystallization algorithm as long carbon chains develop.
For this reason, during Monte-Carlo simulations, we employ reactive empirical bond order (REBO) potential, parameterized for C-H systems.
A comparison of phonon dispersions as obtained from REBO-CH and optimized Tersoff potentials shows that the optimized Tersoff potential reproduces the optical phonon modes more successfully (Fig.~\ref{sfig:pdrs}).

\clearpage
\section{\supplement\ On Vibrational Mode Analysis of MAC}

\subsection{Inverse Participation Ratio}

Amorphous heat carriers are different from phonons.  According to Allen and Feldman, there are different types of amorphous carriers: Extendons (propagons and diffusons) and locons.  Locons are highly localized carriers, while extendons are spatially extended.  One can calculate the inverse participation ratio (IPR), a quantitative analysis of the extent of localization of a vibrational mode, to differentiate locons from extendons and identify the so-called mobility edge. IPR quantifies the degree of spatial localization of a vibrational mode and is defined as
\begin{equation}
	\mathrm{IPR}_m = \sum_i \bigg( \sum_{\alpha} \epsilon_m (i, \alpha) ^2\bigg)^2
\end{equation}
where $\epsilon_m (i, \alpha)$ denotes the $\alpha$th Cartesian components of the normalized polarization vector  associated with the $m$th vibrational mode of the $i$th atom.
An alternative but related measure, the participation ratio (PR), is given by:
\begin{equation}
	\mathrm{PR}_m=     \frac{1}{N} \frac{ \bigg( \sum_i  \sum_{\alpha} \epsilon_m (i, \alpha) ^2\bigg)^2 }{\sum_i \bigg( \sum_{\alpha} \epsilon_m (i, \alpha) ^2\bigg)^2}
\end{equation}
where the notation is consistent with the definition above. The participation ratio is inversely related to the IPR and offers insight into how spatially extended a mode is across the atomic structure.
In perfect crystalline materials, vibrational modes tend to be delocalized, involving coherent motion of many atoms. As a result, their inverse participation ratios are typically low—on the order of
1/N, where
N is the number of atoms. In contrast, disordered or amorphous systems often exhibit localized vibrational modes, for which the IPR approaches unity.

\begin{figure}[h!]

	\includegraphics[height=100mm]{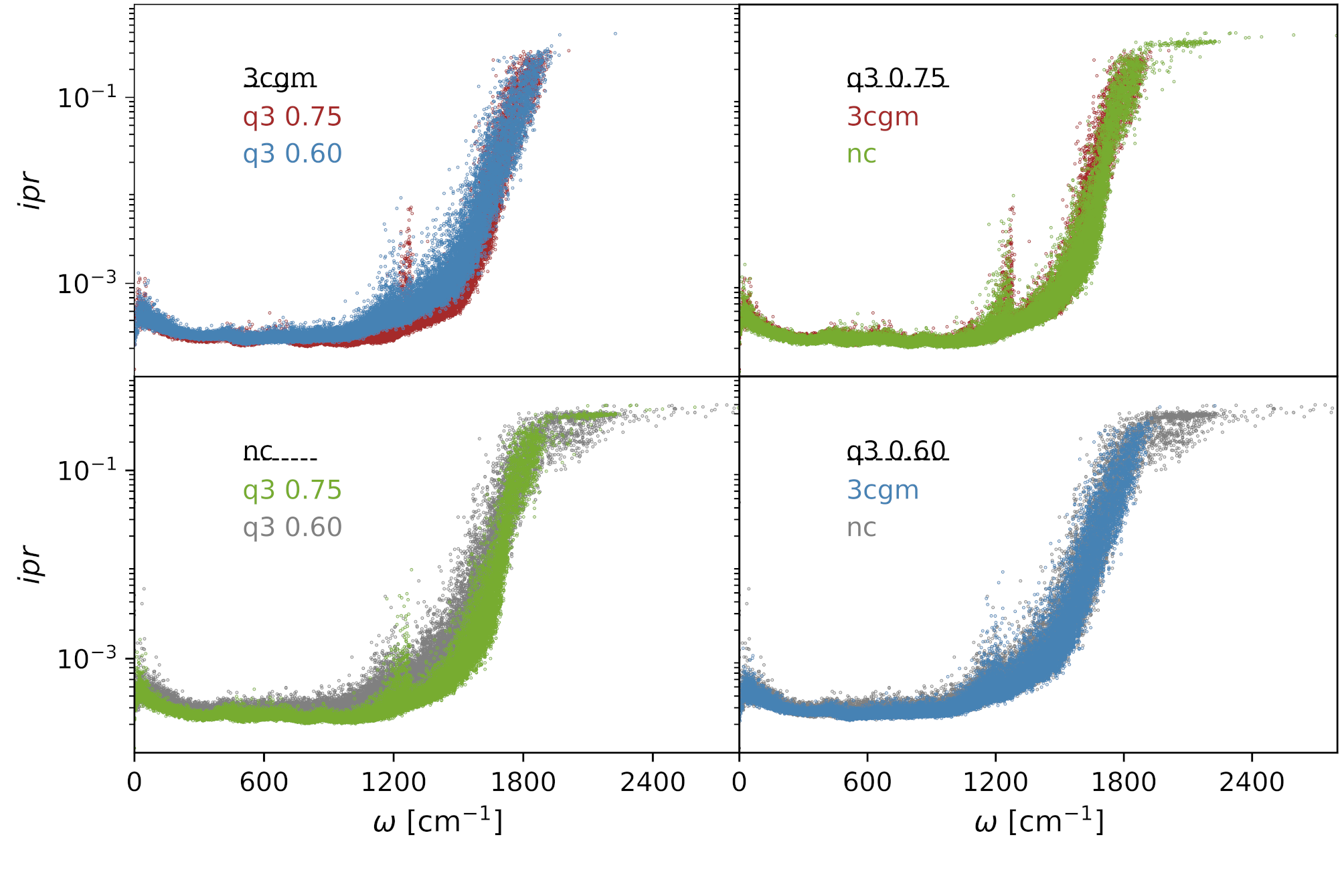}
	\caption{Inverse participation ratios.}
	\label{sfig:siprs}
\end{figure}
All calculated IPRs are printed in Fig.\ref{sfig:siprs}. On the left panels, configurations that belong to the same structural type but different $q_3$ values are compared; on the contrary, on the right panels, configurations that belong to different structural types but the same $q_3$ values are compared. Decreasing $q_3$ results in a higher IPR value, but changing configuration type barely influences the IPR values except for high-energy modes. A prominent localized feature observed in all examined structures occurs near  $1200 $ cm$^{-1}$ , corresponding to the band edge in graphene. The mobility edge—the frequency separating extended and localized modes—is also found around {$1350 $ cm$^{-1}$ when considering the PR$= 0.1$ as the threshold for locons. }

{If the eigenvector is a normalized vector.
\begin{equation*}
	\mathrm{PR}_m=     \frac{1}{N} \frac{1}{\sum_i \bigg( \sum_{\alpha} \epsilon_m (i, \alpha) ^2\bigg)^2}
\end{equation*}
If the eigenvector is not a normalized vector.
\begin{equation*}
	\mathrm{PR}_m=     \frac{1}{N} \frac{ \bigg( \sum_i  \sum_{\alpha} \epsilon_m (i, \alpha) ^2\bigg)^2 }{\sum_i \bigg( \sum_{\alpha} \epsilon_m (i, \alpha) ^2\bigg)^2}
\end{equation*}

\subsection{Dynamical Structure Factor}

Extendons are divided into two subcategories as propagons and diffusons. Propagons are much like phonons; they propagate and have well-defined wave-vectors. Diffusons are also propagating; nevertheless, they do not have meaningful wave-vectors. Instead, they have a diffusive profile. So, we can separate propagons from diffusons by examining the relationship between wave-vector and frequency of the mode.
Fourier transforms,  $F_{\eta}(\boldsymbol{q}, \omega) $, of the eigenmodes' longitudinal, transverse, and out-of-plane components are calculated for this purpose~\cite{beltukov_pre_2016, antidormi-2020-2dmat}.

\begin{align*}
	F_{L}(\boldsymbol{q}, \omega) &= \sum_{i=1}^{3N} \left|\sum_{k=1}^N  \hat{\boldsymbol{q}} \cdot \boldsymbol{\epsilon}_{k}(\omega_i) \; e^{i \boldsymbol{q}\cdot\boldsymbol{R}_k} \right|^2 \delta(\omega - \omega_i) \\
	F_{T}(\boldsymbol{q}, \omega) &= \sum_{i=1}^{3N} \left|\sum_{k=1}^N  \hat{\boldsymbol{q}} \times \boldsymbol{\epsilon}_{k, \parallel}(\omega_i) \; e^{i \boldsymbol{q}\cdot\boldsymbol{R}_k} \right|^2 \delta(\omega - \omega_i)
	\\
	F_{Z}(\boldsymbol{q}, \omega) &= \sum_{i=1}^{3N} \left|\sum_{k=1}^N  \hat{\boldsymbol{q}} \times \boldsymbol{\epsilon}_{k, \perp}(\omega_i) \; e^{i \boldsymbol{q}\cdot\boldsymbol{R}_k} \right|^2 \delta(\omega - \omega_i)
\end{align*}
Here, $\hat{\boldsymbol{q}} = \boldsymbol{q}/|\boldsymbol{q}|$ is the unit vector along $\boldsymbol{q}$. $\boldsymbol{R}_k $ is the position of the $k$th atom. $\boldsymbol{\epsilon}_{k}(\omega_i)$ is the displacement of the $k$th atom  for $i$th mode. $F_{\eta}(\boldsymbol{q}, \omega) $ is  averaged over all possible directions of $\boldsymbol{q}$. During plotting, $F_{\eta}(\boldsymbol{q}, \omega) $ is divided by the magnitude of its maximum for each fixed value of $\omega$ for clarity.

\begin{figure}
	\centering
	\includegraphics[width=120mm]{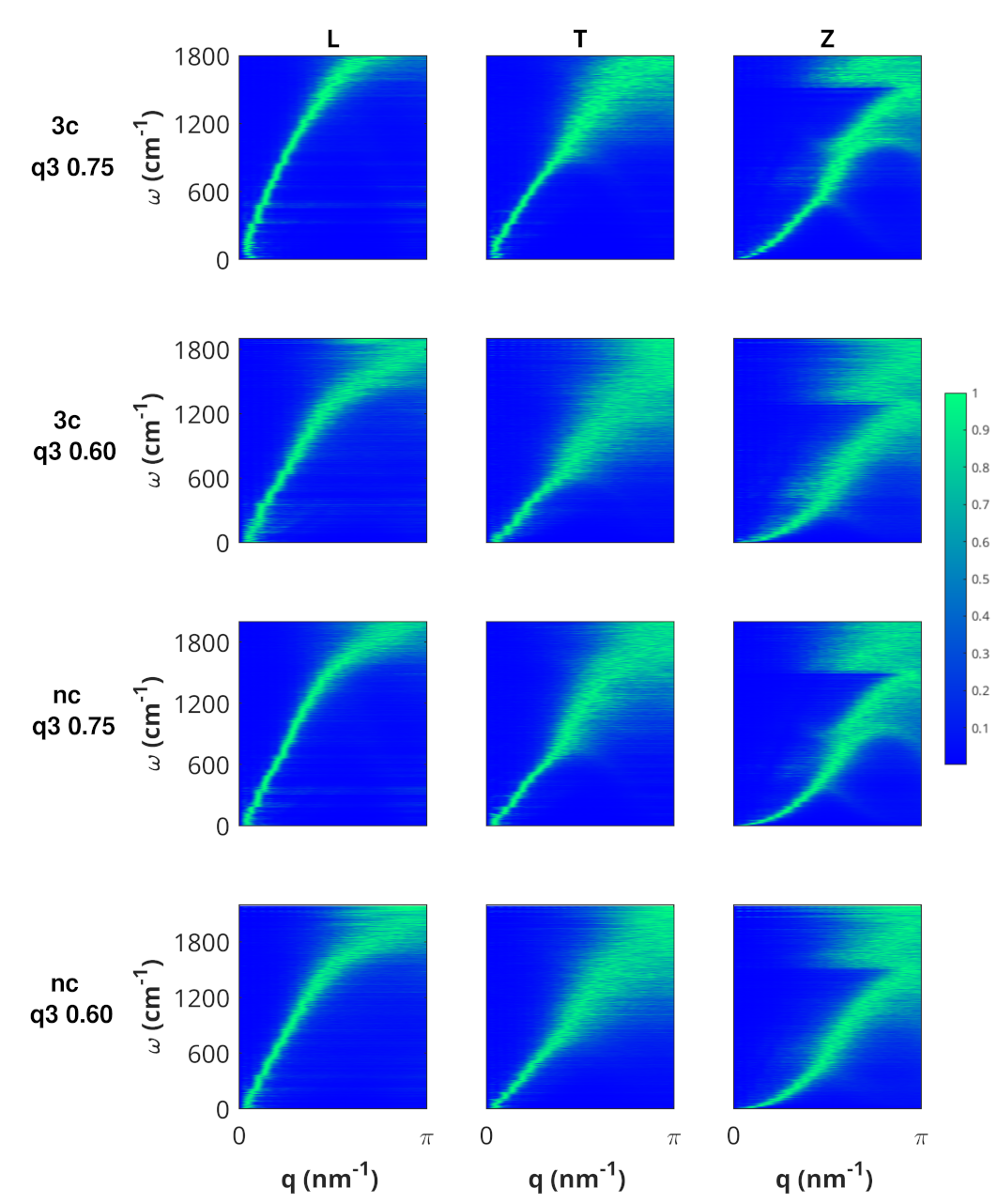}
	\caption{Components of
		eigenmodes in the reciprocal space as a function of the wave number
		$\boldsymbol{q}$ and the frequency $\omega$.}\label{sfig:df}
\end{figure}

In Fig.~\ref{sfig:df}, $F_{\eta}(\boldsymbol{q}, \omega) $ are plotted for $\eta =$ L~(longitudinal), T~(transverse), Z~(out-of-plane) modes. We already know modes with frequency higher than {$1350 \; \text{cm}^{-1}$ }are locons whose dispersion is expected to be scattered.

Only the low-frequency region  ($  0-40 \; \text{cm}^{-1}$)  allows a phonon-like dispersion. The {broadening} starts at lower frequencies when the configuration has a lower $q_3$ value, which is expected since more disorder means more scattering centers and more deviation from phonon-like behavior. The diffusons constitute most of the vibrational spectrum of amorphous monolayer carbon, mainly in the range $40-1200 \; \text{cm}^{-1}$. Our results for dynamical structure factor are consistent with literature, particularly on the Ioffe-Regel limit for amorphous monolayer carbon being around 1~THz (33.34~$\cm$).~\cite{zhu-2016-nanoletters}   Nonetheless, there is a controversy about the Ioffe-Regel limit for amorphous carbon monolayer. Antidormi \emph{et al.} found a much higher Ioffe-Regel limit.~\cite{antidormi-2020-2dmat} For additional insight about the character of the modes,  we refer the reader to the modes' polarization profile as they are discussed in the main text.

\clearpage

\clearpage
\section{Landauer Method for Quantum and Classical Thermal Conductivity}

\subsection{Partitions}

The system is partitioned as shown in Figure~\ref{sfig:GFpartition}.  The central region contains the amorphous structure. Left and right reservoirs need to be free of scatterings and are made of pristine graphene. We include buffer layers between the reservoirs and the central region, which are 8 \AA~wide. They preserve the hexagonal lattice structure, but the atomic positions are relaxed to their equilibrium positions. Periodic boundary conditions are implemented in all directions. The spacing in the out-of-plane direction is large enough to hinder the interlayer interactions.

\begin{figure}[h]
	\vspace{2em}
	\includegraphics[width=.75\textwidth]{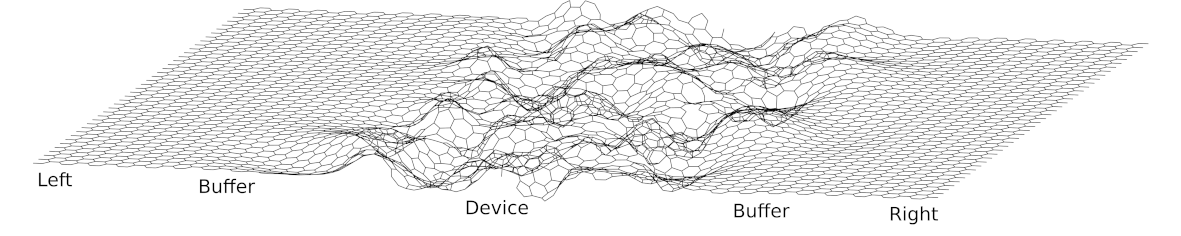}
	\caption{ The partition of the structure as the left, the center, and the right system is shown. The central part includes MAC and buffer regions.}
	\label{sfig:GFpartition}
\end{figure}

\subsection{Green Functions and Thermal Conductivity}

Within the Green’s function formalism, the transmission amplitude is expressed as
\begin{equation}
	\trans(\omega)=
	\mathrm{Tr}
	\left[
	\Gamma^L \mathcal{G}^R \Gamma^R \mathcal{G}^A
	\right],
\end{equation}
where $\Gamma^{L/R}$ denote the reservoir broadening matrices, and $\mathcal{G}^{R/A}$ are the retarded and advanced Green’s functions of the central region coupled to the reservoirs.
The dimensionless weight function entering the Landauer conductance formula is given by
\begin{equation}
	p(\omega,T)=
	\frac{\hbar\omega}{k_B}
	\frac{\partial f_\mathrm{BE}}{\partial T},
\end{equation}
with $f_\mathrm{BE}$ being the Bose–Einstein distribution function.
This function determines how different vibrational modes contribute at a given temperature, and its frequency dependence at varying $T$ is shown in Fig.~\ref{fig:trans_cond}(a).
Together, these expressions fully specify the quantum-mechanical calculation of vibrational thermal conductance and its conversion to thermal conductivity,
\begin{eqnarray}
	\kappa(T)=
	\frac{\kB L}{2\pi A}
	\int_{0}^{\infty}
	d\omega\,
	p(\omega,T)\,
	\trans(\omega),
\end{eqnarray}
with $L$ being the length of the amorphous region and $A$ is the cross section area.

\subsection{Landauer formalism at the classical limit}
In the classical limit, mode populations are determined by the equipartition of energy rather than Bose–Einstein statistics,
\begin{equation}
	f_{cl} = \frac{k_BT}{\hbar\omega}.
\end{equation}
The corresponding weight function becomes
\begin{equation}
	p_{cl}(\omega,T)=
	\frac{\hbar\omega}{k_B}
	\frac{\partial f_{cl}}{\partial T}
	= 1,
\end{equation}
leading to the expression for the thermal conductivity in the elastic limit,
\begin{equation}
	\kappa_{cl}(T)=
	\frac{\kB L}{2\pi A}
	\int d\omega\,
	\trans(\omega).
\end{equation}
This approximation is well justified for amorphous materials, where elastic scattering dominates.
The resulting expression constitutes a classical version of the Landauer approach and enables a direct assessment of quantum versus classical effects in vibrational heat transport.

\clearpage
\subsection{Validation of Size Independence}

\begin{figure}[b]
	\centering
	\begin{tabular}{@{}c@{}}
		\includegraphics[height= 0.25\textheight]{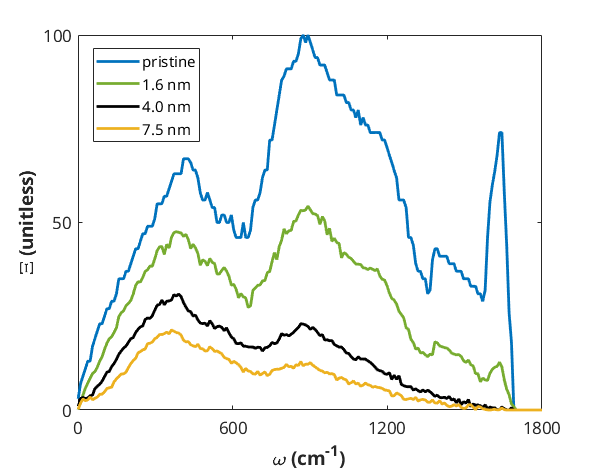} \\[\abovecaptionskip]
		\small (a) $q_3= 0.70$.
	\end{tabular}

	\vspace{\floatsep}

	\begin{tabular}{@{}c@{}}
		\includegraphics[height = 0.25\textheight]{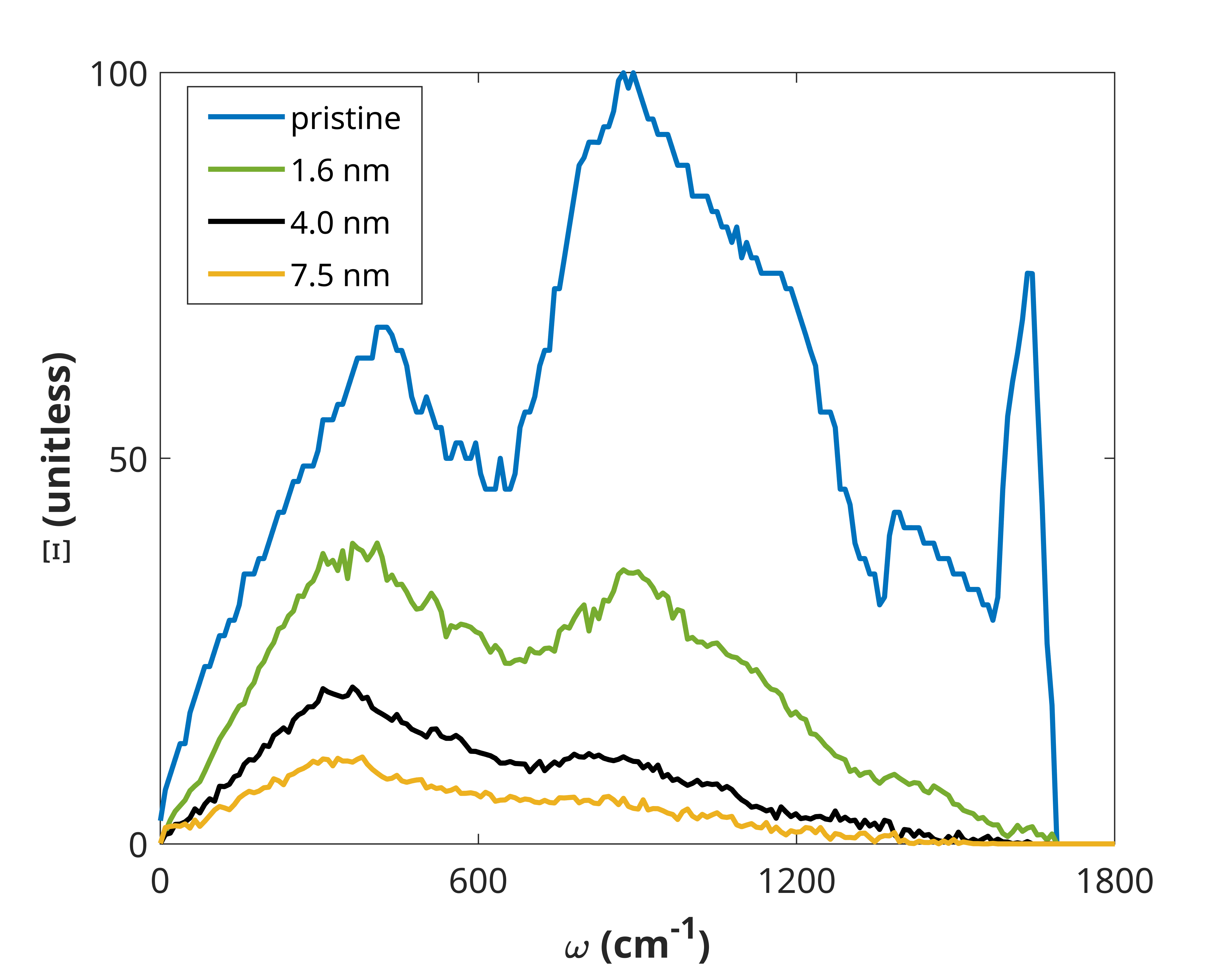} \\[\abovecaptionskip]
		\small (b) $q_3= 0.55$.
	\end{tabular}
	\vspace{\floatsep}

	\caption{Transmission figures for systems with different device lengths and the
		same $q_3$ parameter.}\label{sfig:sizetrans}

\end{figure}

The results presented in the main text correspond to the amorphous regions that are 4~nm in the transport direction.
To check the size effect, we generate 3C configurations with $q_3=0.55$ and 0.70 having lengths between 1.6~nm and 7.5~nm. The total length of the corresponding short (long) system is 11.6 (17.5)~nm. The transverse lengths of all systems are 10~nm. The structures are shown in Fig.~\ref{sfig:short} and Fig.\ref{sfig:long} and their transmission spectra are plotted in Fig.~\ref{sfig:sizetrans}.

Increasing the length of the system lowers the transmission values regarding the $q_3$ value, especially in the high frequency range, since modes with low frequencies have longer MFPs.
Conductivities are shown in Fig.\ref{sfig:sc} using the same color code.
The solid (dashed) curves represent quantum mechanical (classical) thermal conductivities, while the circled (uncircled) lines stand for $q_3$ value of 0.55 (0.70).
Slight differences in the conductivity values are mainly because of differences in atomic details of the configurations, and they are averaged out at higher temperatures, hence at the classical limit.
These results show that the reported conductivity values are largely independent of system size.

\begin{figure}[h!]
	\centering
    \includegraphics[width=0.4\linewidth]{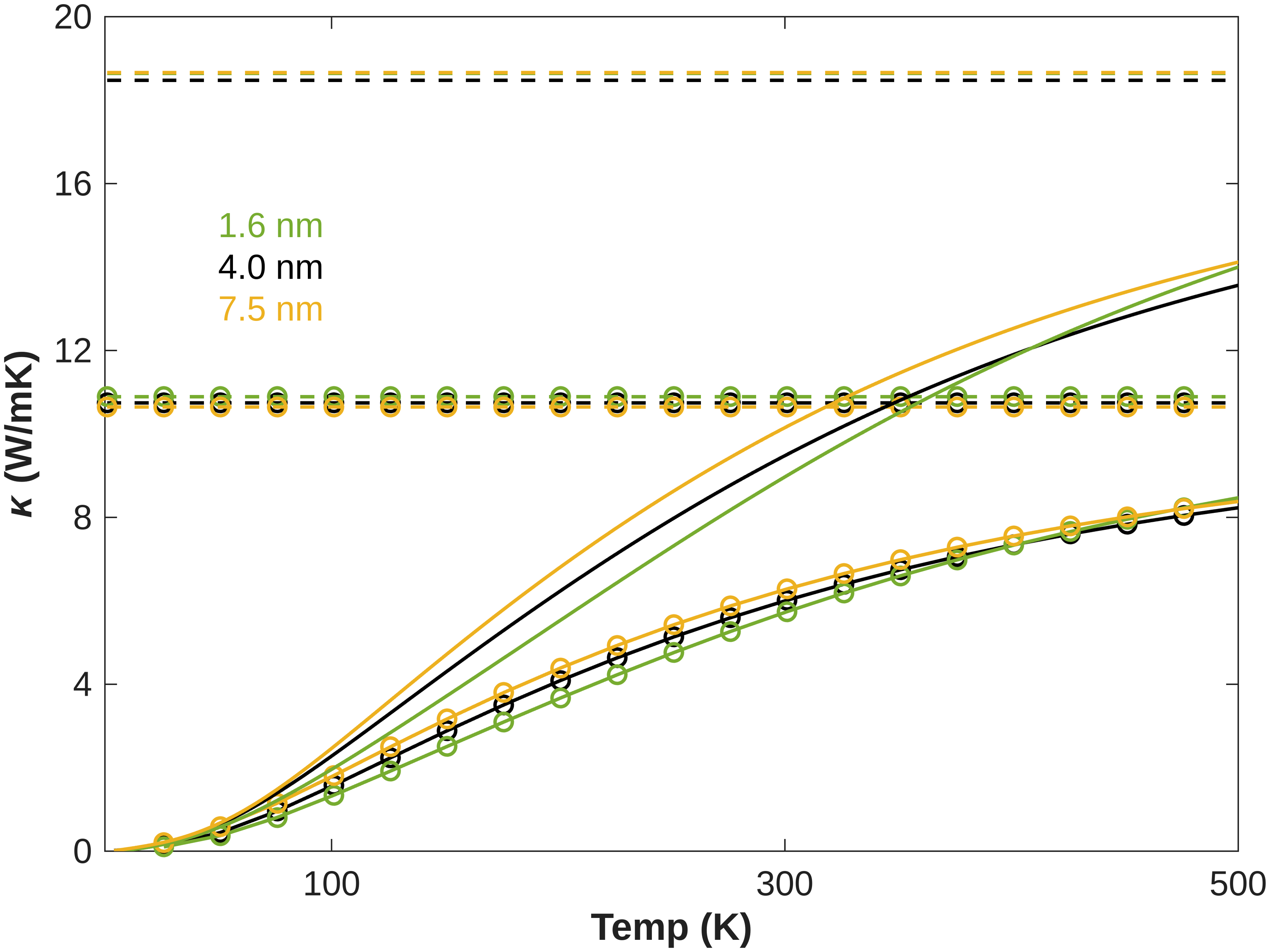}
	\caption{Thermal conductivity values as a function of temperature for systems with different device lengths.}\label{sfig:sc}
\end{figure}

\clearpage
\section{Periodic Boundary Conditions and Kubo-Greenwood Formalism}

In addition to Green's function technique, which involves fictitious partitioning of the system, we also calculated the thermal conductivity using the Kubo-Greenwood (KG) method\cite{li-prb-2011, torres-book-2014}, which is free from size and interface issues.
The system under study is 3C-type configuration with $q_3=0.70$, and has dimensions of $25 \times 25 \; \text{nm}^2$, and fulfills periodic boundary conditions. Based on the Green's function results, we estimate the average phonon mean free path (MFP) for the $q_3=0.70$ configurations to be less than 5~nm. Therefore, our sample size of $25\times25$~nm$^2$ is sufficiently large to rule out finite-size effects.
The calculated MFPs are presented in Fig.~\ref{sfig:agmfp}, with the inset showing the corresponding phonon
transmission values.

MFP values obtained from the KG method are consistent with our earlier estimation using the GF method.
For configurations with $q_3=0.40$, we similarly estimate an average MFP of $\simeq$1~nm, suggesting that this approach provides a reliable approximation.
Further insight comes from examining the transmission spectra for the $q_3=0.70$ configurations, see Fig.~\ref{sfig:sizetrans}(a). Notably, the transmission for a structure of length 7.5~nm is reduced to approximately one-fifth of the ideal (defect-free) value. The inset illustrates even stronger suppression of transmission. When increasing the device length from 1.6~nm to 4.0~nm, the transmission drops significantly more than in the range from 4.0~nm to 7.5~nm. This drop indicates that the device with a 1.6~nm length lies well below the mean free path.

TC, as obtained from KG method, is presented in Fig.~\ref{sfig:conds}, where it is directly compared against results from the GF method with $q_3=0.70$.
We observe that quantum TC obtained from KG and GF methods are in quite good agreement.
Considering the remarkable agreement besides the key differences between the two methods, we conclude that the partitioning scheme, the boundary conditions and the chosen system sizes have minor effects on our results.

\begin{figure}[b]
	\centering
	\includegraphics[width=0.4\textwidth]{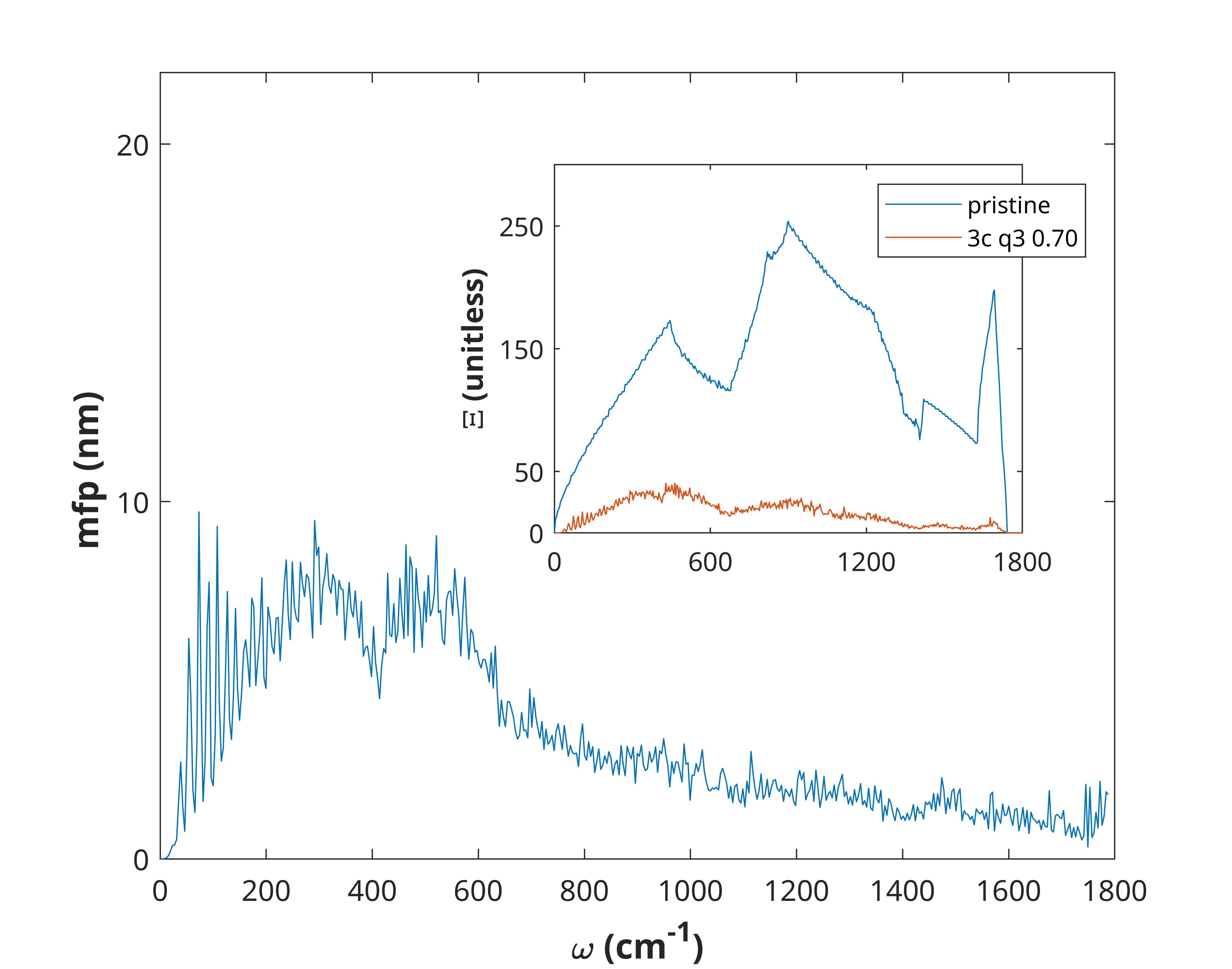}
	\caption{Mean free paths of 3C configuration with $q_3=0.70$. The system size is $25\times25\;\text{nm}^2$. The inset demonstrates the transmissions of pristine and amorphous phases.}\label{sfig:agmfp}
	\includegraphics[width=0.4\linewidth]{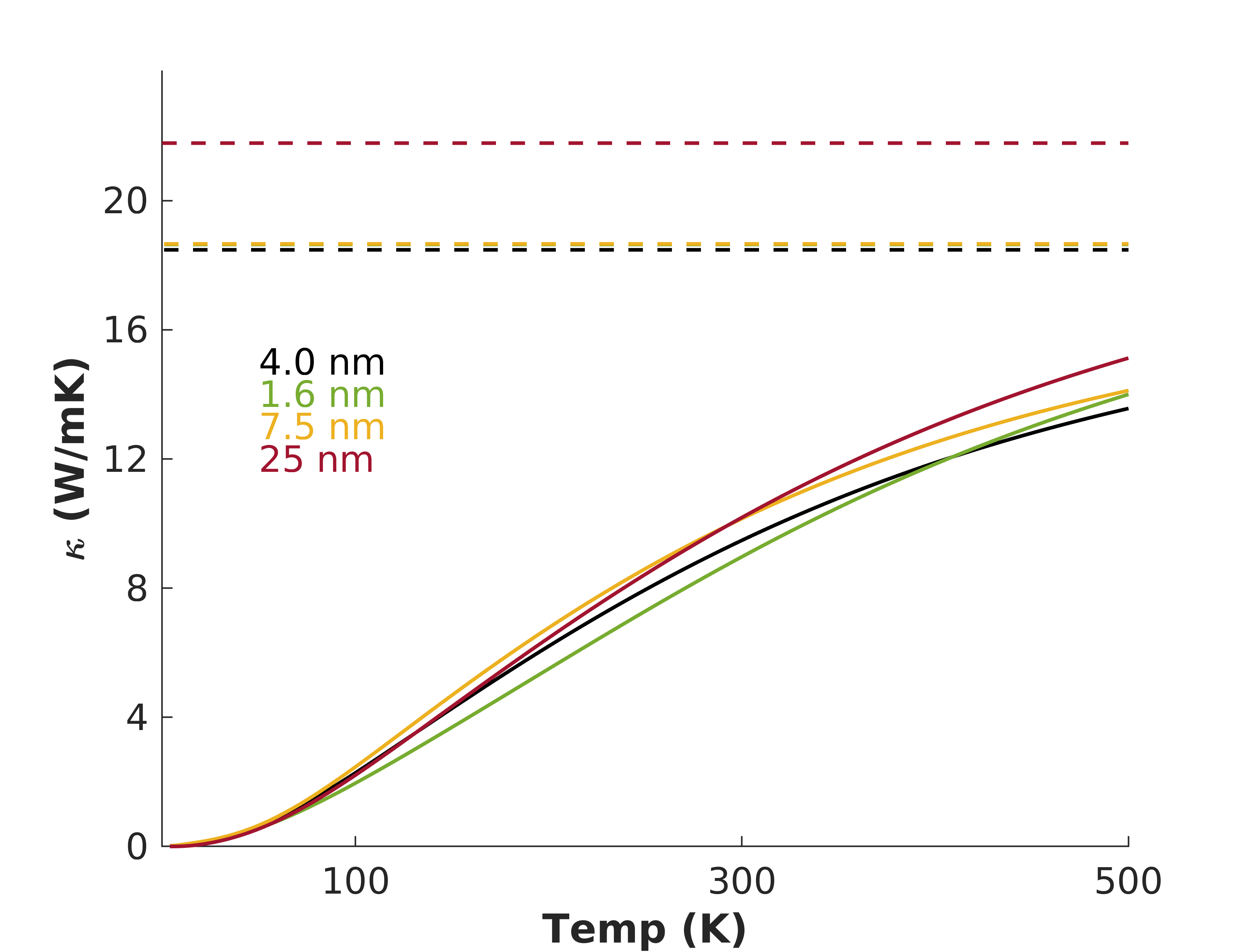}
	\caption{Thermal Conductivities of configurations with $q_3=0.70$. The dashed(solid) lines demonstrate the classical (quantum) conductivities. The 25~nm long configuration is calculated with the Kubo-Greenwood method, while the Green's functions method is applied to others.}\label{sfig:conds}
\end{figure}

\end{document}